\documentclass[twoside,12pt]{article}
\usepackage{hyperref}
\usepackage{bookmark}
\usepackage{xcolor}
\usepackage{indentfirst}
\usepackage{bm}
\usepackage{epsfig}
\usepackage{graphicx}
\usepackage{amsmath}
\usepackage{amssymb}
\usepackage{array}
\usepackage{amsfonts}
\usepackage{braket}
\usepackage{multirow}
\usepackage{textcomp}
\usepackage{manyfoot}
\usepackage{booktabs}
\usepackage{algorithm}
\usepackage{algorithmicx}
\usepackage{algpseudocode}
\usepackage[title]{appendix}
\usepackage[font=small,labelfont=bf]{caption}

\DeclareCaptionLabelFormat{mylabel}{#1 #2.\hspace{1.5ex}}
\DeclareCaptionLabelFormat{mylabel1}{#1 #2.\hspace{1.5ex}}
\newcommand{\bea}{\begin{eqnarray}}
\newcommand{\eea}{\end{eqnarray}}
\newcommand{\be}{\begin{equation}}
\newcommand{\ee}{\end{equation}}

\begin{document}
	\def\Journal#1#2#3#4#5{{\it #1} {\bf #2}\ifx#3\empty\else(#3)\fi, #4 (#5)}
	\def\RPP{{Rep. Prog. Phys}}
	\def\PRC{{Phys. Rev. C}}
	\def\PRD{{Phys. Rev. D}}
	\def\PRB{{Phys. Rev. B}}
	\def\PRA{{Phys. Rev. A}}
	\def\ZPA{{Z. Phys. A}}
	\def\NPA{{Nucl. Phys. A}}
	\def\NPB{{Nucl. Phys. B}}
	\def\JPG{{J. Phys. G }}
	\def\PRL{{Phys. Rev. Lett.}}
	\def\PR{{Phys. Rep.}}
	\def\PLB{{Phys. Lett. B}}
	\def\AP{{Annals Phys.}}
	\def\EPJA{{Eur. Phys. J. A}}
	\def\EPJC{{Eur. Phys. J. C}}
	\def\NP{{Nucl. Phys.}}  
	\def\RMP{{Rev. Mod. Phys.}}
	\def\IJMPE{{Int. J. Mod. Phys. E}}
	\def\IJMPD{{Int. J. Mod. Phys. D}}
	\def\AJ{{Astrophys. J.}}
	\def\AJL{{Astrophys. J. Lett.}}
	\def\AA{{Astron. Astrophys.}}
	\def\ARAA{{Annu. Rev. Astron. Astrophys.}}
	\def\MPLA{{Mod. Phys. Lett. A}}
	\def\ARNPS{{Annu. Rev. Nucl. Part. Sci.}}
	\def\LRR{{Living. Rev. Rel.}}
	\def\CQG{{Class. Quant. Grav.}}
	\def\RAS{{Mon. Not. R. Astron. Soc.}}
	\def\JPA{{J.Phys. A}}
	\def\ATMP{{Adv. Theor. Math. Phys}}
	\def\GRG{{Gen. Rel. Grav.}}
	\def\JPCS{{J. Phys. Conf. Ser.}}
	\def\PRR{{Phys. Rev. Res.}}
	\def\NA{{Nature Astron.}}
	\def\S{{Science}}
	\def\N{{Nature}}
	\def\JCAP{{JCAP}}
    \def\Ph{{Physics}}
    \def\PDD{{Phys. Dark Univ.}}
    \def\JHEP{{JHEP}}
    \def\EPL{{EPL}}
    \def\PPNP{{Prog. Part. Nucl. Phys.}}

\thispagestyle{empty} \vspace*{0.8cm}\hbox
to\textwidth{\vbox{\hfill\huge\sf Commun. Theor. Phys.\hfill}}
\par\noindent\rule[3mm]{\textwidth}{0.2pt}\hspace*{-\textwidth}\noindent
\rule[2.5mm]{\textwidth}{0.2pt}


\begin{center}
\LARGE\bf {A Quantum-Gravity-Motivated GUP Effective Metric}
\end{center}

\footnotetext{\hspace*{-.45cm}\footnotesize $^\dag$Corresponding author, E-mail:  anto.sulaksono@sci.ui.ac.id}

\begin{center}
\rm M. H. Al Ghifari$^{\rm a)}$, \ \ M. F. Fauzi$^{\rm a)}$, \ \ A. Rohim$^{\rm a)}$, \ \ H. S. Ramadhan$^{\rm a)}$, \ \ H. Alatas$^{\rm b)}$ \ and  \ A. Sulaksono$^{\rm a)\dagger}$
\end{center}

\begin{center}
\begin{footnotesize} \sl
${}^{\rm a)}$ Departemen Fisika, FMIPA, Universitas Indonesia, Depok, 16424, Indonesia \\
${}^{\rm b)}$ Theoretical Physics Division, Department of Physics, IPB University, Bogor, 16680, Indonesia
\end{footnotesize}
\end{center}

\begin{center}
\footnotesize (Received XXXX; revised manuscript received XXXX)

\end{center}

\vspace*{2mm}

\begin{center}
\begin{minipage}{15.5cm}
\parindent 20pt\footnotesize
Recent critiques have addressed certain aspects of the generalized uncertainty principle (GUP) effective metric \cite{Ong:2023jkp}. This study presents a scale-dependent quadratic GUP effective metric, constructed through analyses of the gravity-induced phase shift (COW experiment) and the Einstein-Bohr photon box Gedanken experiment. In contrast to the procedure outlined in \cite{Li:2016yfd}, the momentum-dependent metric is improved by introducing an interpolating function $\Delta p (r)$, which employs the effective distance concept to accurately capture the distinct behavior of $\Delta p (r)$ in both short and long distance regimes. The resulting effective metric exhibits the same structure as that derived from the Renormalization Group (RG) theory. However, the RG parameter $\hat{\gamma}$ can now be related to the dimensionless GUP parameter $\beta_0$, thereby distinguishing this metric from the RG-based approach. The corresponding effective metric prediction demonstrates internal consistency of the model, and phenomenologically the predictions are in agreement with some quantum black hole models in some limits. The effective metric improved black hole thermodynamics and shadow predictions compared to the heuristic approach and other proposed GUP effective metrics. Furthermore, the relationship between GUP and $f(R)$ gravity \cite{DAgostino:2025axy} may provide a possible future route toward a more fundamental description of GUP.
\end{minipage}
\end{center}

\begin{center}
\begin{minipage}{15.5cm}
\begin{minipage}[t]{2.3cm}{\bf Keywords:}\end{minipage}
\begin{minipage}[t]{13.1cm}
GUP; Effective distance; Black holes; Quantum gravity
\end{minipage}\par\vglue8pt

\end{minipage}
\end{center}

\section{Introduction}\label{intro}	
Formulating a consistent quantum gravity (QG) theory remains a central challenge in modern physics. Several promising candidates have emerged, including string theory \cite{Kaku:1974zz, Kaku:1974xu}, loop quantum gravity (LQG) \cite{Rovelli:1991zi, Rovelli:1997yv}, gravity based on the renormalization-group (RG) \cite{Bonanno:2000ep, Bonanno:2001xi, Reuter:2003ca}, and noncommutative geometry approaches\cite{Girelli:2004md}. Ref. \cite{Addazi:2021xuf} provides a comprehensive review of recent progress in QG phenomenology. Many QG theories predict a minimum measurable length around the Planck length $ \ell_p$, defined as the Schwarzschild radius of a black hole with the same energy as a photon of that wavelength. Through dimensional analysis, $ \ell_p \equiv \sqrt{\hslash G/c^3}=1.6 \times 10^{-35}$ m, where $c$ is the speed of light, $G$ is the gravitational constant, and $\hslash$ is the Planck constant. This minimal length arising from quantum-gravitational effects alters black hole geometry, eliminates curvature singularities, and leads to the concept of remnants: black holes that do not fully evaporate but approach a minimal mass. Consequently, thermodynamic analysis reveals that the temperature, heat capacity, and luminosity vanish at this mass. In addition, the temperature attains a maximum for a nonzero black hole mass, leading to a phase transition in the heat capacity. This link between minimal length and black hole properties prompts theoretical modifications of the Heisenberg uncertainty principle (HUP). The generalized uncertainty principle (GUP) provides a widely studied phenomenological framework for such modifications; its connection to the running of Newton's constant is described in Ref.\cite{Xiang:2013sza}, and recent developments in GUP are reviewed in Ref.\cite{Bosso:2023aht}. The discussion of effective metrics in GUP will be addressed in detail later.

The second property is that on this scale, the parameters defining the physical model, such as Newton's constant, the cosmological constant, etc., become scale-dependent quantities \cite{Bonanno:2000ep, Bonanno:2001xi, Reuter:2003ca}. The scale dependence is a generic feature of quantum field theory. From a gravity perspective, the scale dependence is expected to modify the horizon. Therefore, it has an impact on the thermodynamics and shadows of BH. See the discussion in~\cite{Lambiase:2022xde} and the references for details.  The main idea of asymptotically safe gravity theory \cite{Bonanno:2000ep, Bonanno:2001xi, Reuter:2003ca,Lambiase:2022xde} is to integrate the differential equation of the renormalization group (RG) for gravitational coupling in the order of Newton's constant $G$ to become a function of the momentum scale $k$, $G(k)$. The key point is the link between the renormalization scale $k$ and the radial coordinate $r$, which can be established in this framework through the concept of a proper distance scale $d(P)$, which provides the relevant cutoff for the Newton constant when the test particle is located at the point $P$ of the object's spacetime. For a spherically symmetric spacetime, it implies that $d(P)$ depends only on the coordinate $r$ of $P$, $d = d(r)$ \cite{Bonanno:2001xi}. Therefore, a complete form of an improved metric solution can be obtained. For BH's case, these RG procedures are expected to modify the classical solutions by incorporating quantum features through a momentum-dependent Newton constant or running Newton constant. One of the most fascinating implications of this procedure is that the BH has a free singularity in the center. Recently, some efforts have been made to study the connection between GUP and asymptotically safe gravity theory \cite{Lambiase:2022xde,Lambiase:2023hng}. However, the corresponding connection is still indirect.  We note that a scale-dependent metric is also found in the Rainbow Gravity theory~\cite{Magueijo:2002xx}, where this theory suggests that spacetime geometry can also depend on the energy of the test particle.
We note that the relation between GUP and the quantum Raychaudhuri is studied in Refs.~\cite{Vagenas:2017fwa,Ali:2015tva}. In contrast, the connection between GUP and Lorentz violation in the standard model extension (SME) is studied in~\cite{Lambiase:2017adh}. 

One issue with the GUP approach is the lack of a  procedure for deriving the GUP's effective metric from fundamental principles. According to the finding in~\cite{AlGhifari:2025xrt}, an appropriate GUP metric is essential to correctly predict the valid range of $\beta$ in the neutron star case. Quite recently, the author of Ref.~\cite{Ong:2023jkp} studied some of the effective metric existence proposed by GUP in the literature~\cite{Scardigli:2014qka,Vagenas:2017vsw,Contreras:2016xib,FaragAli:2015boi,Anacleto:2020lel,Anacleto:2021qoe}. They have found that some of these results are incorrect because the constructions rely too much on heuristic arguments and lack a guiding principle in constructing the effective metric~\cite{Ong:2023jkp}. One of the solutions offered by the author is to start from the expression of entropy with the standard logarithmic correction term and to use the recently proposed "generalized entropy and varying-G" (GEVAG)~\cite{Lu:2024ppa} to obtain the associated metric~\cite{Ong:2025ent}. 

The second issue discussed in~\cite{Bosso:2023aht} is that although much effort has gone into constraining the phenomenological KMM (Kempf-Mangano-Mann) GUP parameter, not all bounds in the literature use consistent standards of rigor. The strongest bound, $\beta < 10^{33}$, comes from scanning tunneling microscope studies but was established more than a decade ago. This underscores the need for new strategies. In particular, adopting an accurate, effective metric for GUP rather than relying on heuristic arguments may yield more reliable constraints from gravitational observables.

To this end, there is an interesting GUP's effective metric proposal~\cite{Saha:2013kta,Li:2016yfd,Farahani:2020ctt} based on the analysis of the Collela, Overhauser, and Werner (COW) experiment~\cite{Colella:1975dq} and Einstein-Bohr's photon box Gedanken experiment, which is overlooked in the discussion in Ref.~\cite{Ong:2025ent}. The interesting fact of this approach is that the effective metric GUP describes a family of spacetimes with different momenta $\Delta p$ scales. This fact is similar to that predicted by rainbow gravity theory and RG-based theories.  Furthermore, an important step forward has been taken by the authors of Ref.~\cite{Li:2016yfd} by analyzing the requirement of $\Delta p$ using tidal force such that 
\be
{(\Delta p)}^2 \geq \frac{\hbar^2}{c^2} \biggl(\frac{2GM}{r^3}\biggr).
\label{abra}
\ee 
This finding is remarkable because, with this relation, we obtain the complete form of the fully radial dependence of the GUP effective metric based on this procedure. Furthermore, for the KMM GUP model case~\cite{Kempf:1994su}, where $\hat{z}(\hat{p}) \equiv 1 + \frac{\beta_0 \ell_p^2}{\hbar^2} \hat{p}^2$, we can obtain an effective metric similar to that predicted by the Hayward regular BH model \cite{Hayward:2005gi,Cadoni:2022chn}, which is non-singular in the center. We expect that the effective metric of the KMM GUP model obtained by this procedure yields predictions for the horizon similar to those of the regular BH models. If we expand to the region with $r \gg \ell_p$, the effective metric coincides with the one predicted by a model of the LQG approach~\cite{Lewandowski:2022zce}. The latter means that we can constrain the dimensionless deformation parameter $\beta_0$ of the KMM GUP model~\cite{Kempf:1994su} of LQG~\cite{Lewandowski:2022zce}. It is also worth noting that Eq.~(\ref{abra}) is similar to the proper distance $d(r)$ at the limit $r \rightarrow 0$ (UV region) obtained in Ref.~\cite{Bonanno:2000ep} i.e.,  
\be
\lim_{r \rightarrow 0} k^2 \approx \frac{9}{4} \zeta^2 \frac{\hbar^2}{c^2} \biggl(\frac{2GM}{r^3}\biggr),
\label{der}
\ee 
where $\zeta$ is a numerical constant to be fixed. However, for the limit $r \rightarrow \infty$ (IR region), the authors of~\cite{Bonanno:2000ep} obtain a different expression, i.e.,
\be
\lim_{r \rightarrow \infty} k^2 \approx \zeta^2/r^2.
\label{der2}
\ee

Inspired by~\cite{Bonanno:2000ep}, which uses the effective distance concept, we expect that Eq.~(\ref{abra}) can be improved by considering the different behavior of the momentum scale $\Delta p$ in large and small regions $r$. Building on this, we will also examine whether our approach addresses the concerns raised in~\cite{Ong:2023jkp} regarding the GUP's effective metric, specifically the existence of a remnant at zero temperature and zero entropy, as well as the phase transition in the black hole heat capacity. These issues are central to our work.

First, this work investigates the GUP's effective metric studied in~\cite{Li:2016yfd}. However, because the proper distance exhibits different behavior at large and small distances, for the $\Delta p \longleftrightarrow r$ relation, instead of using Eq. \eqref{abra}, we use an interpolation inspired by the concept of effective distance of Bonnano-Reuter in a form of effective parameterization function~\cite{Bonanno:2000ep} where in this way, we may improve the IR part of the GUP effective metric proposed in ~\cite{Li:2016yfd}. The relation is as follows:
\be
{(\Delta p)}^2 \geq \zeta^2 \frac{\hbar^2}{c^2} \frac{(r+ \hat{\gamma} GM)}{r^3}, 
\label{abraX}
\ee 
where $\zeta$ and $\hat{\gamma}$ are parameters that we will discuss further in the next section. It can be seen that if $\zeta$ is set to be 1 and $\hat{\gamma}$ is set to be 2, Eq. \eqref{abraX} reduces back to Eq. \eqref{abra} for small $r$ and becomes $1/r^2$ for large $r$. We emphasize that this distinct behavior serves as a crucial indicator for addressing the  collapse of GR in both the UV and the IR regions. Second, we try to constrain $\beta_0$ using an effective metric predicted by one of the existing QG proposals. Third, we study the BH thermodynamics and possible observational tests, such as the shadow predictions of this proposal, and compare the results with those of other quantum BH models. 

Here, we construct a scale-dependent GUP-effective metric with a minimal length, which is internally consistent and captures several commonly discussed quantum-gravity-inspired black-hole features. In contrast to the heuristic GUP approach and the GUP effective metric studied in \cite{Ong:2023jkp}, our results are phenomenologically in agreement with other quantum BH results in some limits, including the quantum Raychaudhuri equation, and yield acceptable quantum BH thermodynamics and shadow predictions. The reason that the heuristic GUP does not properly predict remnants in entropy and Hawking temperature is that the metric's behavior in the UV region is not properly accounted for. We build a metric based on the heuristic GUP using this formula. GUP is often overlooked as a pathway to quantum gravity phenomenology in the semi-classical regime because it lacks fundamentality, as it is unconnected to the action. The procedure in \cite{DAgostino:2025axy} established the connection between GUP deformation and curvature corrections, opening the way to finding the generalized action of GUP in the future.

We have organized this paper as follows: Section \hyperref[sec:GUP]{2} introduces the GUP model. Section \hyperref[sec:EM]{3} examines the effective metric. Section \hyperref[sec:TP]{4} discusses its impact on BH thermodynamics. Section \hyperref[sec:SP]{5} considers the effect of the effective GUP metric on observational tests. Section \hyperref[sec:FR]{6} discusses the connection between GUP and $f(R)$ gravity. Section \hyperref[sec:Conclu]{7} summarizes our conclusions.

\section{Generalized Uncertainty Principle}\label{sec:GUP}
This section briefly discusses the GUP model used in this work, namely the quadratic GUP model. In this model, the effect of minimal length in quantum mechanics is a deformation of the Heisenberg algebra into quadratic functions of the momentum operator~\cite{Kempf:1994su}.

The three-dimensional commutator of this model has the following form \cite{Kempf:1994su}:
\begin{equation}
	\left[\hat{x}_i,\hat{p}_j\right]=\mathrm{i}\delta_{ij}\hslash\left(1+\frac{\beta}{\hslash^2} \hat{p}^2\right).
	\label{HCR}
\end{equation}
However,  the operators of position $x_i$ and momentum $p_i$ still satisfy the following commutation relations:
\begin{eqnarray}
	[\hat{x}_i,\hat{x}_j]=0; ~~~~ [\hat{p}_i,\hat{p}_j]=0.
\end{eqnarray}
Here, $\hat{p}^2$ is the magnitude of the 3-vector momentum operator and $\beta$ is the phenomenological parameter to accommodate the deformation due to GUP. This parameter can be written as $\beta = \beta_0 \ell_p^2 = \beta_0/M_p^2$. Thus, $\beta_0$ is a dimensionless parameter. Based on the modified commutator in Eq. (\ref{HCR}) and the fact that $\braket{\hat{p}}=0$, the uncertainty relation between position $\Delta x$ and momentum $\Delta p$ is recast in the following form
\begin{equation}\label{P1}
    \Delta x\Delta p\geq\frac{\hslash}{2}\left[1+\frac{\beta}{\hslash^2} \left(\Delta p\right)^2\right].
\end{equation}
From the above inequality, one can obtain a lower bound on the measurable distance uncertainty as:
\begin{equation}
	\Delta x_{\rm min}\approx   \sqrt{\beta}=\ell_p\sqrt{\beta_0}.
	\label{The Minimum x}
\end{equation}
As a non-vanishing minimal uncertainty in position, Eq. (\ref{The Minimum x}) implies the existence of a minimal uncertainty length only given by $\beta_0 >$ 0 or a positive value of  $\beta$.

\section{Black-hole effective metric}\label{sec:EM}
  
Here, we briefly outline the main idea for constructing an effective metric by analyzing the gravity-induced phase shift from Colella, Overhauser, and Werner(COW) and the Einstein-Bohr photon box gedanken experiments. Please, see Refs.~\cite {Saha:2013kta,Li:2016yfd,Farahani:2020ctt} for details. 
Before that, let us write Eq. \eqref{HCR} in general form as 
\be
[\hat{x},\hat{p}]= \mathrm{i} \hbar \hat{z}(\hat{p}),
\label{GUP}
\ee 
and we denote
\be
z\equiv \braket{\hat{z}}.
\ee 

The effective metric proposal of the GUP model is constructed here on the basis of the following concepts.
\begin{itemize} 
	\item Modified wave-particle duality: The uncertainty principle is related to the wave-particle duality. The Broglie formula is expected to be modified when the Heisenberg commutators are modified. It can be shown that the Broglie wave function, for the GUP, obtains the following relation~\cite{Li:2016yfd}

\be
\frac{\mathrm{d}}{\mathrm{d}p}\left(\frac{2 \pi}{\lambda}\right)=\hbar^{-1} z^{-1},
\label{add1}
\ee 
$\lambda$ is the Broglie wavelength. In consequence, there is a common eigenstate $\psi_p$ of eigenvalue $ p$ and $k(p)$ the Broglie's wave number in GUP, which satisfies the corresponding eigen equations. Here~\cite{Li:2016yfd} 
\be
k(p)=\hbar^{-1} \int z^{-1}(p) \mathrm{d}p. 
\label{add2}
\ee
The modified wave-particle is characterized by Eqs.~\eqref{add1} and~\eqref{add2}, which are used as a basis for the following conclusions. 
\item Gravity-Induced quantum phase shift:  COW observed the gravity-induced quantum interference pattern of two neutron beams. When the plane of the two beams is vertical and horizontal, the phase shift is given by~\cite{Li:2016yfd}
\be
\Delta \phi=\frac{mgA}{\hbar v},  
\ee
where $m$ is the mass of the neutrons, $g$ denotes the gravitational field of the earth, $v$ is the velocity of the neutrons. $A=yl$, where y is the height of the upper horizontal path with respect to the lower horizontal one, while $l$ is the length of each horizontal path. Note that the schematic of the experimental setup can be seen in Fig. 1 of Ref.~\cite{Li:2016yfd}. Using Eqs.~\eqref{add1} and  Eqs.~\eqref{add2} it can also be deduced the phase shift for the GUP's case, ie~\cite{Li:2016yfd}
\be
\Delta \phi'=\frac{mg'A}{\hbar v},  
\ee
with $g'=g/z$. This means that, in the GUP case, the COW experiment tells us that two neutron beams propagating in an effective gravitational field $g'$ instead of $g$. 

\item Einstein-Bohr's photon box: This gendanken experiment for weighing a photon, where the correct result can only be obtained by including the time dilation effect. In this experiment, for the clock in the box, the time uncertainty within the HUP due to vertical uncertainty $\Delta x$ is given by~\cite{Li:2016yfd}
\be
\Delta t=\frac{g \Delta x}{c^2} t,
\ee 
Using a similar argument for the GUP case, we can obtain a similar time uncertainty but with relation again  $g$ replaced by $g'$, or we can reach the same conclusion as in the COW experiment~\cite{Li:2016yfd} that the time difference of the photon propagation in the GUP based on this experiment is proportional to $g'$ instead of $g$. Therefore, from these results, it can be concluded that $G'=G/z$.  
\end{itemize}

Based on both experiments, we can conclude that  the consistency of the Newton coupling constant with the Heisenberg uncertainty modification in Eq. \eqref{GUP} is only obtained  when 
\bea
G \Rightarrow G'\left(\Delta p\right)=\frac{G}{z \left(\Delta p\right)}.
\eea
The Newton coupling constant depends on the momentum scale $\Delta p$ and the GUP model used through $z$. Therefore, we can restate what was mentioned in~\cite{Li:2016yfd} that "GUP and running constant $G'$ are two sides of the same coin."

When $G$ is replaced by $G'$, we obtain a modified Schwarzschild metric as follows:
\be
 \mathrm{d}s^2 = -A(r)\mathrm{d}t^2+A(r)^{-1}\mathrm{d}r^2+r^2\mathrm{d}\Omega^2, \label{GUP-Modified SCH}
\ee
with $A(r)$, in general, can be written as
\be
A(r)=\left(1-\frac{2G'\left(\Delta p\right) M}{c^2 r}\right).
\label{general metric tensor}
\ee
This metric is characterized by the effective Newton constant in the sense of modification of Heisenberg's principle and the usual quantum theory. It describes a family of spacetimes that depend on different momentum scales. This behavior is similar to the RG-based gravity and rainbow gravity approaches~\cite{Li:2016yfd, Bonanno:2001xi, Magueijo:2002xx}. However, unlike the case of the RG-based theory, for GUP there is not much study of UV and IR scale identification or momentum cutoffs to relate $\Delta p$ to a function of $r$. Only Ref. ~\cite{Xiang:2013sza} studies that in the UV region, They deduced the identification from gravitation tidal force. 
 On the other hand, although there have been many works on scale identification in  RG-based theory for quantum BH, the choice of effective distance $d(r)$ that determines the $ r$ dependence of the momentum scale is not unique because there is a lack of a physical guiding principle that must be satisfied when making the identification. Please, see Refs. ~\cite{Pawlowski2018,Ishibashi2021,Chen2022} and the references therein for detailed discussions.   
 Therefore, in this work, we borrow the form of effective parameterization function of scale identification in the RG theory of Bonanno Reuter~\cite{Bonanno:2000ep} applied to the GUP phenomenological model, as discussed in the introduction. Note that the choice of an effective parameterization of the scale identification is not unique. However, at least this choice has been tested and the physical predictions are quite acceptable\cite{Bonanno:2000ep}.
 
 It means, unlike the one performed in~\cite{Li:2016yfd}, which used Eq.~\eqref{abra}, here we use an effective parameterization function in Eq.~\eqref{abraX} to obtain the complete form of the radial dependence of the corresponding effective metric, and the result is 
\begin{align}
    A_{TW}(r)=1-\frac{2GMr^2}{r^3+\beta_0\ell_p^2\zeta^2(r+\hat{\gamma} GM)}, \label{Bonanno-Reuter-GUP Metric 2}
\end{align}
where the subscript $TW$ stands for "This Work". We use the unit $c = \hbar = 1$ to simplify the expressions below. It is obvious that the behavior of the metric in Eq. ~\eqref{Bonanno-Reuter-GUP Metric 2} close to the center is different from all GUP's effective metrics discussed in~\cite{Ong:2023jkp}, namely, regular in the center. Considering $A_{TW}(r)=1-2G'M/r$, one can easily identify the running Newton constant from Eq.~\eqref{Bonanno-Reuter-GUP Metric 2} above as
\begin{align}
    G'=\frac{Gr^3}{r^3+\beta_0\ell_p^2\zeta^2(r+\hat{\gamma} GM)}. \label{Running Newton constants}
\end{align}

Note that $\zeta$ is merely a scaling factor that can be absorbed without loss of generality. Therefore, the number of free parameters in this model is two, $\beta_0\zeta^2$ and $\hat{\gamma}$.
Remarkably, the effective Newton constant of quadratic GUP takes the same form as the RG approach~\cite{Bonanno:2000ep}. From the above discussion, it is clear that even though the form is similar to RG-improved geometry, the construction of $G(\Delta p)$ is different because, for the GUP metric obtained based on COW and photon box gedanken experiments instead of the RG equation.  

Furthermore, if we take $\hat{\gamma}=2$, $\zeta=1$, and consider $r\ll\hat{\gamma}GM$, Eq.~\eqref{Bonanno-Reuter-GUP Metric 2} reduces to that $A(r)$ proposed in~\cite{Li:2016yfd}:
\begin{align}
    A_{TW}(r)\approx A_{Xi}(r)=1-\frac{2GMr^2}{r^3+2\beta_0\ell_p^2GM}. \label{Xiang metric}
\end{align}
This reduction shows that our proposed metric provides the complete radial dependence we have mentioned before, which is absent in ~\cite{Li:2016yfd}. For $\ell_p\ll r\ll\hat{\gamma} GM$, we can expand Eq.~\eqref{Bonanno-Reuter-GUP Metric 2} as
\begin{align}
    A_{TW}(r)\approx1-\frac{2GM}{r}+\frac{2G^2M^2\beta_0\ell_p^2\zeta^2\hat{\gamma}}{r^4}+\mathcal{O}\biggl(\frac{1}{r^3}\biggr), \label{Bonanno-Reuter-GUP for small r}
\end{align}
where the $1/r^3$ term becomes less significant than the $1/r^4$ term. For large $r$, Eq.~\eqref{Bonanno-Reuter-GUP Metric 2} takes the form:
\begin{align}
    A_{TW}(r)\approx1-\frac{2GM}{r}+\frac{2GM\beta_0\ell_p^2\zeta^2}{r^3}+\mathcal{O}\biggl(\frac{1}{r^4}\biggr), \label{Bonanno-Reuter-GUP for large r}
\end{align}
where the $1/r^4$ term becomes less significant than the $1/r^3$ term. The expansion at large $r$ is not predicted in~\cite{Li:2016yfd}, implying that our metric imposes the quantum correction for large and small radius. In addition to $\hat{\gamma}=2$, a recent study took the value of the parameter $\hat{\gamma}$ equal to $9/2$ \cite{Bonanno:2000ep,Lambiase:2022xde}. However, the following argument based on their work allows one to treat the parameter $\hat{\gamma}$ as a free positive parameter because this range shows the same qualitative properties in the improved BH spacetime. This unfixed value of $\hat{\gamma}$ is also stated in the RG based theory \cite{Lambiase:2022xde}. We will determine the fixed value of the GUP parameter $\beta_0$ by comparing our metric with references, so the bound in $\hat{\gamma}$ can also be determined by comparing with the LQG approach in \cite{Lewandowski:2022zce}.

As the first step in constraining $\hat{\gamma}$, let us recall a metric component $A_{BR}(r)$ from the Bonanno and Reuter work \cite{Bonanno:2000ep}:
\begin{align}
    A_{BR}(r)=1-\frac{2GMr^2}{r^3+\tilde{\omega}G[r+\hat{\gamma}GM]}. \label{Metric Bonanno-Reuter}
\end{align}
The coincidence of Eq.~\eqref{Bonanno-Reuter-GUP Metric 2} with Eq.~\eqref{Metric Bonanno-Reuter} is obtained using $\ell_p^2\equiv G$, i.e., the condition obtained by taking $c=\hbar=1$. This coincidence allows us to connect $\beta_0$ and $\tilde{\omega}$ of RG theory~\cite{Bonanno:2001xi} as 
\begin{align}
    \beta_0\zeta^2\equiv \tilde{\omega}. \label{beta-omega relation}
\end{align}

Then, the other parameters $\tilde{\omega}$ and  $\hat{\gamma}$ are determined from the results of the leading-order quantum correction of Newton's potential from the standard perturbative quantization of Einstein gravity~\cite{Donoghue:1993eb} and the LQG result\cite{Lewandowski:2022zce}. For large $r$, Ref.~\cite{Bonanno:2000ep} finds $\tilde{\omega}\equiv 118/5\pi$ using quantum corrections by~\cite{Hamber:1995cq}. This concatenation allows us to constrain the GUP parameter $\beta_0\equiv\frac{118}{5\pi}$. However, the literature remains uncertain about the value and even the sign of the leading-order quantum correction to the Newton potential. Please, see \cite{Lambiase:2022xde,Donoghue:1993eb,Bjerrum-Bohr:2002fji,Hamber:1995cq,Frob:2021mpb,Dalvit:1997yc,dePaulaNetto:2021axj} and the references therein for the corresponding discussions. Therefore, the constraint of $\beta_0$ from the quantum correction in the Newtonian potential results is highly dependent on the accuracy of the quantum correction in the Newtonian potential calculation. Note that the correction of Newton potential is defined as $V(r)=-G'M/r$. Within the running Newton constants in Eq.~\eqref{Running Newton constants}, we can write the quantum correction of the Newton potential using the GUP parameter $\beta_0$ as

\begin{align}
    V(r)=-\frac{GM}{r}\biggl(1-\frac{\beta_0G}{r^2}+...\biggr). \label{Correction of Newton potential}
\end{align}

It is also evident that the expansion of our proposed metric, written in Eq.~\eqref{Bonanno-Reuter-GUP for small r}, shares the same structure as the one obtained by~\cite{Lewandowski:2022zce} using the LQG approach , i.e.,
\begin{align}
    A_{LQG}(r)=1-\frac{2GM}{r}+\frac{16\sqrt{3}\pi\gamma^3G^2\ell_p^2M^2}{r^4}, \label{Lewandwoski metric}
\end{align}
where $\gamma$ in Eq. \eqref{Lewandwoski metric} is the Barbero-Immirzi parameter taken within the range~\cite{Domagala:2004jt}:
\begin{equation}
\frac{\ln2}{\pi}<\gamma<\frac{\ln3}{\pi}. 
\end{equation}
Therefore,  we can constrain $\hat{\gamma}$ from the LQG result by comparing Eq.~\eqref{Bonanno-Reuter-GUP for large r} and \eqref{Lewandwoski metric}. In this way, we obtain $\hat{\gamma}$ as 
\begin{align}
    \hat{\gamma}\equiv\frac{8\sqrt{3}\pi\gamma^3}{\beta_0}. \label{Hat gamma}
\end{align}

From Eq. \eqref{Hat gamma}, it can be seen that the uncertainty of $\hat{\gamma}$ depends on the uncertainty of $\beta_0$ and the Barbero-Immirzi parameter $\gamma$. Within Eq.~\eqref{Hat gamma}, we can obtain some possible constraints $\hat{\gamma}$ as shown in Table~\ref{table 1}. In the third column, we provide the references where the quantum correction in the Newton potential is taken to constrain $\beta_0$. Note $118/15\pi\approx2.5$,  $127/30\pi^2\approx0,429$, and  $167/30\pi\approx1.772$. It is also evident that according to Eq. \eqref{Hat gamma}, in GUP, the effective metric is slightly different from that in the RG theory~\cite{Bonanno:2000ep}, where in the latter theory, $\hat{\gamma}$ and $\tilde{\omega}$ are independent parameters. Table~\ref{table 1} provides three different regions with respect to the magnitude of the relatively large acceptable parameter space due to the constraint uncertainty that might provide different behavior of the BH properties. We will discuss this in the next section.

\begin{table}[h]
\centering
\tabcolsep=15pt
\renewcommand\arraystretch{1.2}
    \caption{The possible values of $\beta_0$ and $\hat{\gamma}$ showed the uncertainty for both parameters.}\label{table 1}
    \begin{tabular}{|c|c|c|}
    \hline
        $\beta_0$ & $\hat{\gamma}$ & Refs.\\
    \hline
        $0.429$ & $1.095\lessapprox\hat{\gamma}\lessapprox4.351$ & \cite{Donoghue:1993eb}\\
        $2.5$ & $0.188\lessapprox\hat{\gamma}\lessapprox0.745$ & \cite{Bonanno:2000ep}\\
        $1.772$ & $0.265\lessapprox\hat{\gamma}\lessapprox1.053$ & \cite{Bjerrum-Bohr:2002fji}\\
    \hline
    \end{tabular}
\end{table}

 Note that we take the constraint with a positive value of $\tilde{\omega}$ \cite{Bonanno:2000ep,Donoghue:1993eb,Bjerrum-Bohr:2002fji}. This is because the positive requirement of $\Delta x_{\mathrm{min}}$ in Eq.~\eqref{The Minimum x} guaranties the existence of a minimal length. As a result, we obtain $\hat{\gamma}$ that is smaller than that used in \cite{Bonanno:2000ep}, namely, it is in the range $0.188\lessapprox\hat{\gamma}\lessapprox0.745$. Furthermore, we note that recent work on leading order quantum corrections to the Newtonian potential prefers a negative prediction of $\beta_0$ \cite{Lambiase:2022xde,Frob:2021mpb,Dalvit:1997yc,dePaulaNetto:2021axj}. Nevertheless, other approaches or BH's observations might shed light on this sign issue of $\beta_0$. 

\begin{center}
\begin{figure*}
    \centering
    \includegraphics[scale=0.65]{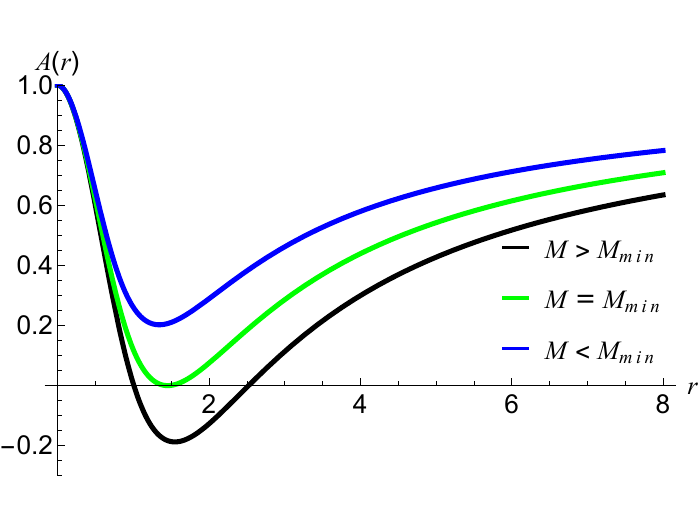}
    \caption{ The GUP-improved BH metric as a function of $r$. The calculation is performed using $\hat{\gamma}=2$ and $\beta_0=0.5$.\label{fig:1}}
\end{figure*}
\end{center}

\begin{figure*}[ht!]
    \centering
    \includegraphics[scale=0.65]{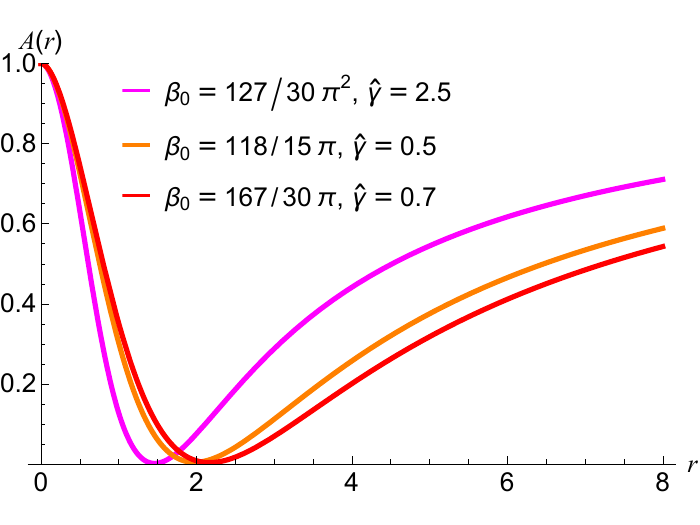}
    \caption{The GUP-improved BH metric as a function of $r$ with various $\hat{\gamma}$ and $\beta_0$.}
	\label{fig:2}       
\end{figure*}

The quantum correction in our metric, imposed by the GUP parameter $\beta_0$, produces a minimal mass $M_{\mathrm{min}}$ that the quantum BH should have, so it has at least one horizon. As shown in Fig.~\ref{fig:1}, at certain fixed $\beta_0$ and $\hat{\gamma}$, the quantum correction will work effectively from $M=M_{\mathrm{min}}$ (the green line). The black line shows the implication of regularity in the center \cite{Cadoni:2022chn}, with BHs potentially having both Cauchy (inner) and event (outer) horizons. Regarding stability, not all Cauchy horizons in regular black hole models are unstable due to mass inflation. Ref.~\cite{Bonanno:2020fgp} demonstrates that the Cauchy horizon of the regular black hole proposed by Hayward and asymptotic-safety is stable due to the suppression of mass inflation. Furthermore, because the structure of this GUP effective metric resembles that of the effective RG theory metric~\cite{Bonanno:2000ep}, we expect the Cauchy horizon of this metric to exhibit stability similarly against the effects of mass inflation. As the mass increases, the quantum correction decreases and approaches GR at a sufficiently large mass (as will be seen clearly in the next section).

Meanwhile, the metric does not exhibit a horizon in the small-mass region $M<M_{\mathrm{min}}$, as indicated by the blue line. This region provides a star solution \cite{Cadoni:2022chn}. This result shows the generic metric behavior of the regular BH solutions. The possibility of a zero, one, or two horizons appearing is generic and independent of the value of $\hat{\gamma}$ and $\beta_0$, as we can see in Fig.~\ref{fig:2}. However, we can also see in Fig.~\ref{fig:2} that the position of the event horizon $r_H$, which has a consequence on the value of the remnant mass $M_{\mathrm{min}}$, is affected by the chosen value of $\hat{\gamma}$ and $\beta_0$.  

For additional comments, it is also worth noting the connection between this metric and the quantum Raychaudhuri equation for null geodesics in~\cite{Ali:2015tva}. Remarkably, the metric in Eqs. \eqref{Bonanno-Reuter-GUP Metric 2} and \eqref{Xiang metric} also obeys the differential equation in Eq. (20) in~\cite{Ali:2015tva}. Therefore, these metrics are also solutions of the quantum Raychaudhuri equation. Recently, interesting work has been done connecting GUP with non-local theories of gravity~\cite{Capozziello:2025iwn} to constrain $\beta_0$. However, they used a different effective metric for quadratic GUP than that proposed in this work. They obtain a negative $\beta_0$. Therefore, it is interesting to estimate the value $\beta_0$ in this way by comparing Eq. \eqref{Correction of Newton potential} with the potential of nonlocal gravity in Eq. (10) in Ref. ~\cite{Capozziello:2025iwn} and to evaluate the characteristic length scale of nonlocality in the ultraviolet limit $L_{UV}$. In this way, we have found a different result that $\beta_0$ is positive, i.e., $\beta_0$ =$[1+ {\rm erf}(1/2)]L_{UV}^2$. We also need to note that from the study of the connection between BH thermodynamics and Landauer's principle in information theory for quantum-improved Schwarzschild BH through RG correction~\cite{Jana:2025hgv}, the authors have found $\tilde{\omega} \ge \frac{\ln 2}{4 \pi} \approx 0.05$. Therefore, we can estimate from this constraint that $\beta_0 \ge 0.05$. Our results are in agreement with this finding.
\section{Black-hole thermodynamics without and with effective metric approach}
\label{sec:TP}
\subsection{Standard heuristic GUP results from effective metric prespective}
We will briefly note the results of the GUP-heuristic correction to BH thermodynamics (see \cite{Adler:2001vs}), which has been criticized in \cite{Ong:2023jkp}. This approach starts with the same uncertainty relation as in Eq.~\eqref{P1} and yields the minimum position uncertainty given by Eq.~\eqref{The Minimum x}. The Hawking temperature is derived heuristically, using the uncertainty principle and the general properties of a black hole. Defining $\Delta x$ as the Schwarzschild radius $R_s$, $\Delta p$ as $1/2\Delta x$, and setting the Boltzmann constant $k_B=1$, the GUP modification in Hawking temperature can be read as \cite{Ong:2023jkp,Ong:2025ent}
\begin{align}
    T_{HGUP1}=\frac{M}{\pi\beta_0}\biggl(1-\sqrt{1-\frac{\beta_0}{4 G M^2}}\biggr). \label{Hawking temperature Ong}
\end{align}

\begin{figure*}[ht!]
	\centering
	\includegraphics[scale=0.45]{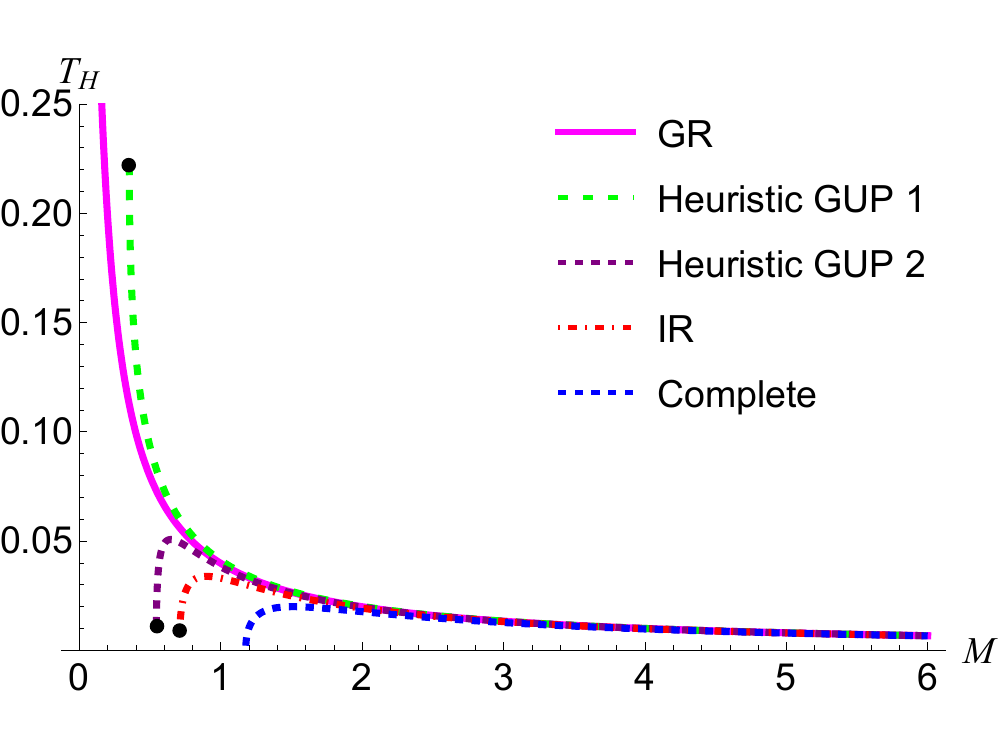}
	\caption{Hawking temperature as a function of mass, where the black dots represent the finite mass. IR part of the effective distance occurs when $\Delta p(r)\geq \zeta/r$,   	while the complete one takes the $\Delta p(r)$ form as Eq.~\eqref{abraX}.}
	\label{justification}       
\end{figure*}

It is worth noting that Eq.~\eqref{Hawking temperature Ong} can also be obtained from an effective metric formalism. We can apply the effective metric procedure discussed in the previous section using~\cite{Ong:2023jkp}
\begin{align}
\Delta p\geq \frac{\hbar}{\beta_0 \ell_p^2}\left[\Delta x\left(1-\sqrt{1-\frac{\beta_0 \ell_p^2}{\Delta x^2}}\right)\right]. \label{momentum uncertainty Ong}
\end{align}
Eq. \eqref{momentum uncertainty Ong} can be considered as scale identification and takes $z$ from Eq. \eqref{P1}, then substitutes into Eq. \eqref{general metric tensor} to obtain $A(r)$, the Hawking temperature can be calculated by using the standard surface gravity equation as
\begin{align}
    T_H = \frac{1}{4\pi} \frac{\mathrm{d}A(r)}{\mathrm{d}r}\bigg|_{r=r_H}. \label{Hawking temperature}
\end{align}
The precise analytic expression of the Hawking temperature can be obtained exactly as in Eq.~\eqref{Hawking temperature Ong} by setting $\Delta x\propto r_H \equiv 2M$ (consider $G=1$) from the beginning. This calculation yields a finite minimal mass with nonzero temperature, as shown in the green line of Fig.~\ref{justification} (Heuristic GUP 1). However, it is obvious that the initial assumption—the radius equals the Schwarzschild horizon radius—is improper. This assumption also leads to an incorrect $\frac{\mathrm{d}A(r)}{\mathrm{d}r}\bigg|_{r=r_H}$.   

The minimum uncertainty of position is related to the minimum mass $M_{min}$ by the standard heuristic-based definition $\Delta x_{\mathrm{min}}= 2GM_{\mathrm{min}} $. One can substitute this relation into Eq.~\eqref{The Minimum x} to obtain the minimum mass as written below:
\begin{align}
    M_{\mathrm{min}}=\frac{\sqrt{\beta_0} M_p}{2 }. \label{minimum mass}
\end{align}
Taking into account the first law of thermodynamics $dS_{HGUP 1}=dM/T_{HGUP 1}$, the entropy can also be obtained as
\bea
    S_{HGUP 1}&=& 2 \pi {\left(\frac{M}{M_p}\right)}^2 +2 \pi \left(\frac{M} {M_p}\right) {\left[{\left(\frac{M}{M_p}\right)}^2-\frac{\beta_0}{4}\right]}^{\frac{1}{2}}\nonumber\\
             &-&\frac{\beta_0 \pi}{2} \ln{\left[ \left(\frac{M}{M_p}\right) + {\left[{\left(\frac{M}{M_p}\right)}^2-\frac{\beta_0}{4}\right]}^{\frac{1}{2}}\right]} + c\nonumber\\
             \label{Ong entropy}
\eea
where $c$ is a constant. For large $M$, Eq. ~\eqref{Ong entropy} becomes
\be
 S_{HGUP 1}= 4 \pi \left({\frac{M} {M_p}}\right)^2 -\frac{1}{2} \pi \beta_0 \ln \left(\frac{M}{M_p}\right) +...\;.
\ee 
The first term is the Bekenstein-Hawking entropy. GUP provides a logarithmic correction with coefficient $-\frac{1}{2} \pi \beta_0$ (negative). In this approach, the horizon radius $r_H$ is assumed to be $r_H=2GM$~\cite{Ong:2023jkp}. This assumption leads to inconsistencies in the calculation of the Hawking temperature, as discussed in~\cite{Ong:2025ent}.

We are intrigued by the failure of the standard heuristic GUP to yield the correct remnant mass, which should be finite as the Hawking temperature drops to zero. The function $\Delta p(r)$, introduced earlier, becomes relevant when we adopt the correct assumption $\Delta x \propto r$. This leads to the same line element structure as in Eq.~\eqref{GUP-Modified SCH}, with the explicit expression given by
\begin{align}
	A_{HGUP 2}(r)=1-\frac{2GM\beta_0\ell_p^4}{\beta_0\ell_p^4r+r^3\left(1-\sqrt{1-\frac{\beta_0\ell_p^2}{r^2}}\right)^2}. \label{metric Ong}
\end{align}
One can easily obtain the standard Schwarzschild form when the GUP effect is neglected by taking $\beta_0\rightarrow0$. By inserting Eq.~\eqref{metric Ong} above into Eq.~\eqref{Hawking temperature}, the heuristic Hawking temperature, shown by the purple line in Fig.~\ref{justification} (Heuristic GUP 2), shares the same behavior as the one that uses the infrared (IR) part of the effective distance. This is shown in the red line in Fig.~\ref{justification}. This similarity arises from the fact that the expansion of $\Delta p(r)$ of Heuristic GUP 2 is proportional to $1/2r$. This matches the form $\Delta p(r)$ takes in the IR when $\zeta\rightarrow 1/2$. The difference in the remnant positions between IR and HGUP2 comes from the added correction after the leading order of the expansion in Eq.~\eqref{momentum uncertainty Ong} for HGUP2. Both cases yield a finite minimum mass that also occurs at a nonzero temperature. However, unlike Heuristic GUP 1, Heuristic GUP 2 has a maximum Hawking temperature that can yield a phase transition in the heat capacity. However, the remnant mass with exactly zero temperature is still absent. The blue line shows the complete form of the effective distance, which accounts for the different behaviors of the effective distance in the IR and UV regions, as $\Delta p(r)$ shows in Eq.~\eqref{abraX}. The effective distance in the UV region is the main contributor to obtaining the correct remnant mass at zero $T_H$.

To this end, Fig.~\ref{justification} demonstrates that the GUP effective metric approximation, which adds the effective distance concept to cover the IR and UV regions, is quantum mechanically justified, improves the standard heuristic approach in the UV region, and properly calculates the Hawking temperature through the surface gravity formula. It is also worth noting that, with the existing correct GUP effective metric, we can extend the application, for example, to properly calculate the properties of horizonless objects such as neutron stars and white dwarfs \cite {AlGhifari:2025xrt}.

\subsection{GUP Effective metric compared to other quantum BH models and constraining GUP parameter}
In the previous section, we discussed the connection between our metric in Eq.~\eqref{Bonanno-Reuter-GUP Metric 2} and those in Refs. \cite{Bonanno:2000ep} and \cite{Lewandowski:2022zce}, obtaining constraints for $\beta_0$ and $\hat{\gamma}$ based on these links. We have also discussed that by choosing $\hat{\gamma}=2$ and $\zeta=1$, which align with the parameter range defined by our constraints, and considering $r\ll\hat{\gamma}GM$, our metric reduces to the form proposed by Ref. \cite{Li:2016yfd}. Here, $\hat{\gamma}$ determines the strength of the quantum gravity corrections, while $\zeta$ serves as a scaling parameter. In this subsection, we investigate the thermodynamic properties of black holes described by our proposed GUP-modified metric—specifically the Hawking temperature, heat capacity, and entropy—and compare our findings with those from Refs. \cite{Li:2016yfd,Bonanno:2000ep,Lewandowski:2022zce}. 

From Eq.~\eqref{Hawking temperature}, we can calculate the Hawking temperature of our proposed metric in Eq.~\eqref{Bonanno-Reuter-GUP Metric 2}:

\begin{align}
    T_{TW}=\frac{\ell_p^2 M_{TW} r_H \left( r_H^3 - \ell_p^2 r_H \beta_0 - 2 \ell_p^4 M_{TW} \beta_0 \hat{\gamma} \right)}{2 \pi \left( r_H^3 + \ell_p^2 r_H \beta_0 + \ell_p^4 M_{TW} \beta_0 \hat{\gamma} \right)^2}. \label{Our Hawking temperature}
\end{align}
The Hawking temperature given in Eq.~\eqref{Our Hawking temperature} reduces to the standard GR Hawking temperature $T_{GR}=1/8\pi GM$ when the GUP effect vanishes ($\beta_0\rightarrow0$) or for large $r_H$. Thus, quantum effects are important mainly for small-mass black holes. Section~\ref{sec:EM} notes that the metric in Ref.~\cite{Li:2016yfd} is a reduced form of Eq.~\eqref{Bonanno-Reuter-GUP Metric 2}, leading to Eq.~\eqref{Xiang metric}. The corresponding Hawking temperature is
\begin{align}
    T_{Xi}=\frac{\ell_p^2 M_{Xi} r_H \left( r_H^3 - 4 \ell_p^4 M_{Xi} \beta_0 \right)}{2 \pi \left( r_H^3 + 2 \ell_p^4 M_{Xi} \beta_0 \right)^2}. \label{Xiang Hawking temperature}
\end{align}
One can take $\hat{\gamma}=2$ and $r\ll\ell_p^2r_H\beta_0$ to find that Eq.~\eqref{Our Hawking temperature} is reduced to Eq.~\eqref{Xiang Hawking temperature}, and also Eq.~\eqref{Our black hole mass} to Eq.~\eqref{Xiang black hole mass}. We also compare our result with Ref.~\cite{Bonanno:2000ep} and Ref.~\cite{Lewandowski:2022zce} where the Hawking temperature based on those metrics (Eq.~\eqref{Metric Bonanno-Reuter} and Eq.~\eqref{Lewandwoski metric}) is
\begin{align}
    T_{BR}&=\frac{\ell_p^2 M_{BR} r_H \left( r_H^3 - \ell_p^2r_H\tilde{\omega} - 2 \ell_p^4 M_{BR} \tilde{\omega}\hat{\gamma}   \right)}{2 \pi \left( r_H^3 + G \tilde{\omega} (r_H + G M_{BR} \hat{\gamma}) \right)^2}, \label{Bonanno-Reuter Hawking temperature} \\
    T_{LQG}&=\frac{\ell_p^2M_{LQG} r_H(r_H^3-32\sqrt{3}\ell_p^4M_{LQG} \pi\gamma^3)}{2\pi r_H^6}. \label{Lewandowski Hawking temperature}
\end{align}

To study how BH thermodynamic properties evolve as the BH becomes larger, we need to find the BH mass as a function of the radius of the outer horizon. We do this by setting $A(r)=0$ in Eq.~\eqref{GUP-Modified SCH}. Repeating this for Eqs.~\eqref{Bonanno-Reuter-GUP Metric 2}, \eqref{Xiang metric}, \eqref{Metric Bonanno-Reuter}, and \eqref{Lewandwoski metric} gives $M(r_H)$ for our metric, Ref.~\cite{Li:2016yfd}, Ref.~\cite{Bonanno:2000ep}, and Ref.~\cite{Lewandowski:2022zce}.

\begin{align}
    M_{TW}&=\frac{r_H^3+\ell_p^2r_H\beta_0}{2\ell_p^2r_H^2-\hat{\gamma}\ell_p^4\beta_0}, \label{Our black hole mass} \\
    M_{Xi}&=\frac{r_H^3}{2\ell_p^2r_H^2-2\ell_p^4\beta_0}, \label{Xiang black hole mass} \\
    M_{BR}&=\frac{r_H^3+\ell_p^2r_H\tilde{\omega}}{2\ell_p^2r_H^2-\hat{\gamma}\ell_p^4\tilde{\omega}}, \label{Bonanno-Reuter black hole mass} \\
    M_{LQG}&=\frac{r_H^3}{\ell_p^2r_H^2+r_H\sqrt{\ell_p^4\left(r_H^2-16\sqrt{3}\ell_p^2\pi\gamma^3\right)}}. \label{Lewandowski black hole mass}
\end{align}

Note that for a regular black hole, there is a positive minimum mass, i.e., remnant mass. Therefore, the right-handed parts of Eq. \eqref{Our black hole mass} for parameters shown in Table \ref{table 2} as well as the ones from Eqs.  \eqref{Xiang black hole mass}-\eqref{Lewandowski black hole mass} and the corresponding Hawking temperatures are always positive and never diverge due to $r_H\gg l_p$.  The BH mass has a model-dependent relation to $r_H$. Thus, all thermodynamic quantities, such as Hawking temperature, entropy, and heat capacity, depend on the BH mass or the radius of the horizon. The heuristic approach criticized in Ref.~\cite{Ong:2023jkp} does not explicitly provide BH mass information because its effective metric is absent from the Hawking temperature calculation. Therefore, we show the evolution of Hawking temperature and entropy—Eqs.~\eqref{Hawking temperature Ong} and \eqref{Ong entropy}—for the heuristic approach only as a function of the BH mass in Figs.~\ref{fig:4} and ~\ref{fig:6}.

We define three cases to study how GUP affects the properties of BH (see Table~\ref{table 2}). Note that in the third column we provide the references where the quantum correction in the Newton potential is taken to constrain $\hat{\gamma}$. In Case 1, we fix $\beta_0$ from~\cite{Bonanno:2000ep} as discussed in Table \ref{table 1}, and consider one value within the range of the corresponding $\hat{\gamma}$, i.e., $\hat{\gamma}=0.5$. In case 2, we fix $\hat{\gamma}=2$, since at this value, our metric can coincide with the metric in~\cite{Li:2016yfd}. The bound on $\beta_0$ is obtained by substituting this $\hat{\gamma}$ into Eq.~\eqref{Hat gamma}: $0.235\lessapprox\beta_0\lessapprox0.933$, and we take it from the range $\beta_0=0.5$. For case 3, we fix $\hat{\gamma}=9/2$, i.e, the value used in \cite{Bonanno:2000ep,Lambiase:2022xde}. From this $\hat{\gamma}$, we obtain $0.587\lessapprox\beta_0\lessapprox2.333$, and then we take $\beta_0=1.5$ as our consideration. 

\begin{table}
\centering
\tabcolsep=15pt
\renewcommand\arraystretch{1.2}
    \caption{The cases to study how GUP affects BH properties. }\label{table 2}
    \begin{tabular}{|c|c|c|c|}
    \hline
    {Case} & {$\beta_0$} & {$\hat{\gamma}$} & {Refs.}\\
    \hline
    {1} & {$2.5$} & {$0.5$} & {\cite{Bonanno:2000ep}}\\
    {2} & {$0.5$} & {2} & {\cite{Li:2016yfd}}\\
    {3} & {$1.5$} & {9/2} & {\cite{Bonanno:2000ep}}\\
    \hline
    \end{tabular}
\end{table}

\begin{figure*}[ht!]
    \centering
    \includegraphics[scale=0.65]{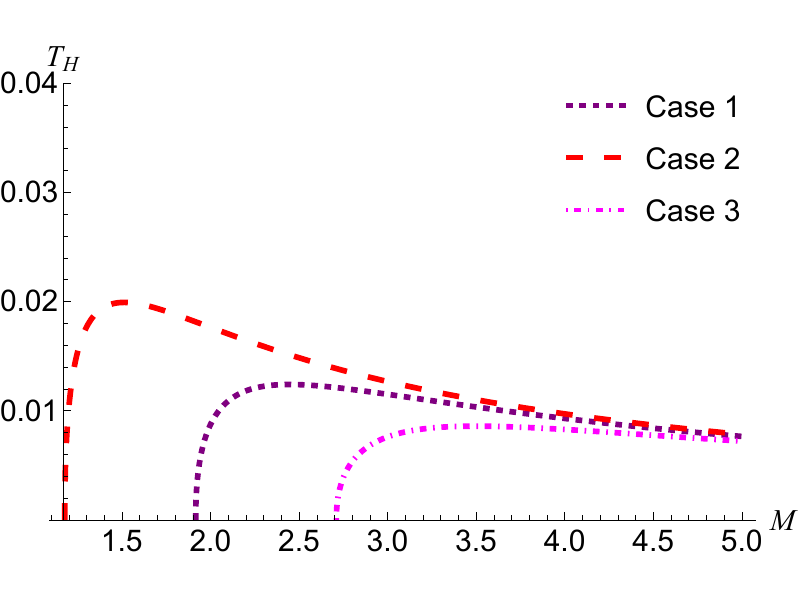}
    \caption{Hawking temperature as a function of mass with 3 cases $\beta_0$ and $\hat{\gamma}$ variation as defined in Table~\ref{table 2}.}
    \label{fig:3}       
\end{figure*}

\begin{figure*}[ht!]
    \centering
    \includegraphics[scale=0.5]{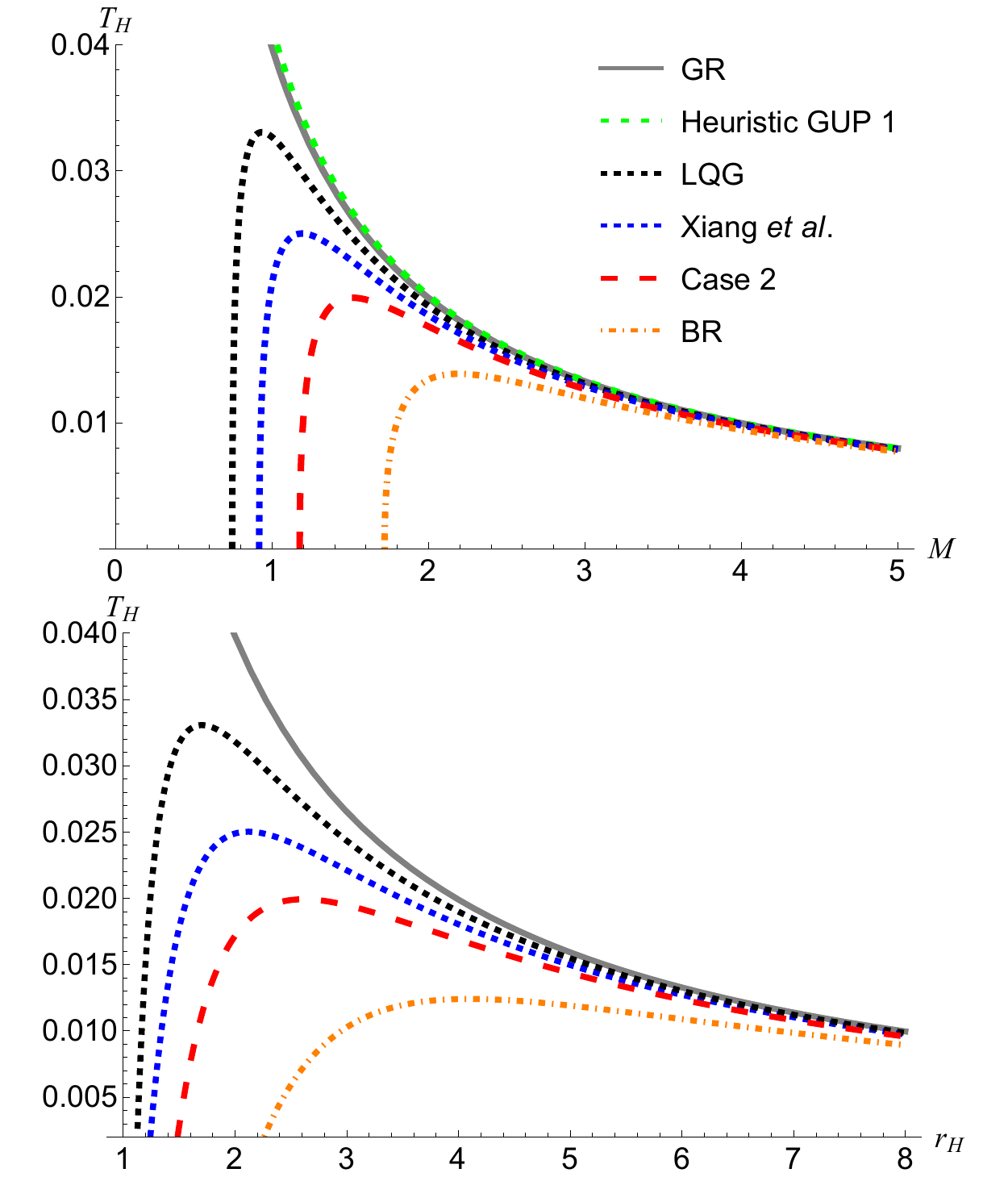}
    \caption{Hawking temperature as a function of mass and horizon radius.}
    \label{fig:4}       
\end{figure*}

In the previous section, we already discussed how $\beta_0$ and $\hat{\gamma}$ shift the minimum mass in the metric plot as a function of radius (Fig.~\ref{fig:2}). The impact of these parameters on the minimum mass shift due to Hawking temperature is shown in Fig.~\ref{fig:3}. Figure~\ref{fig:4} compares our results with those from other models. Specifically, the top panel of Fig.~\ref{fig:4} depicts the relationship between Hawking temperature and BH mass, while the bottom panel illustrates Hawking temperature versus horizon radius. In both panels, note that the reference models provide baseline trends for comparison with our metric. From Eq.~\eqref{Hawking temperature Ong}, under the heuristic GUP, obtaining a minimum mass at a nonzero temperature implies no remnant mass—a feature we previously mentioned and now observe more distinctly in Fig.~\ref{fig:4}. For our metric's Hawking-temperature analysis, we plot only Case 2 in Fig.~\ref{fig:4} to exemplify this behavior, since Cases 1 and 3 show a similar trend, as seen in Fig.~\ref{fig:3}. In particular, the values of $\hat{\gamma}$ and $\beta_0$ in Case 2 are set to match those of the plot by \cite{Li:2016yfd}, allowing for a direct comparison. However, despite using the same $\hat{\gamma}$ and $\beta_0$, our metric—with its additional terms—produces a distinct Hawking temperature profile at small $r$ compared to Ref.~\cite{Li:2016yfd}.

We need to discuss the Barbero-Immirzi parameter $\gamma$, which is particularly related to the LQG metric in \cite{Lewandowski:2022zce}. From loop quantum cosmology in Ref.~\cite{Barboza:2022hng}, there are several possibilities of $\gamma$, including a smaller region than what we prefer for our metric. For our Hawking temperature in Eq.~\eqref{Our Hawking temperature}, small $\gamma$ leads to small $\hat{\gamma}$, as in Eq.~\eqref{Hat gamma}. There are no other consequences as $\hat{\gamma}$ becomes small, in addition to the shift of the minimum mass as shown in Fig.~\ref{fig:3}. However, in the context of the metric in \cite{Lewandowski:2022zce}, as we can see from Eq.~\eqref{Lewandowski Hawking temperature}, there is a minimum $\gamma$ that we can take to maintain the existence of a remnant mass of its metric. It can be seen that although all models exhibit different trends in the small $r_H$ and $M$ regions, they coincide in the large $r_H$ and $M$.

\begin{figure*}[ht!]
    \centering
    \includegraphics[scale=0.5]{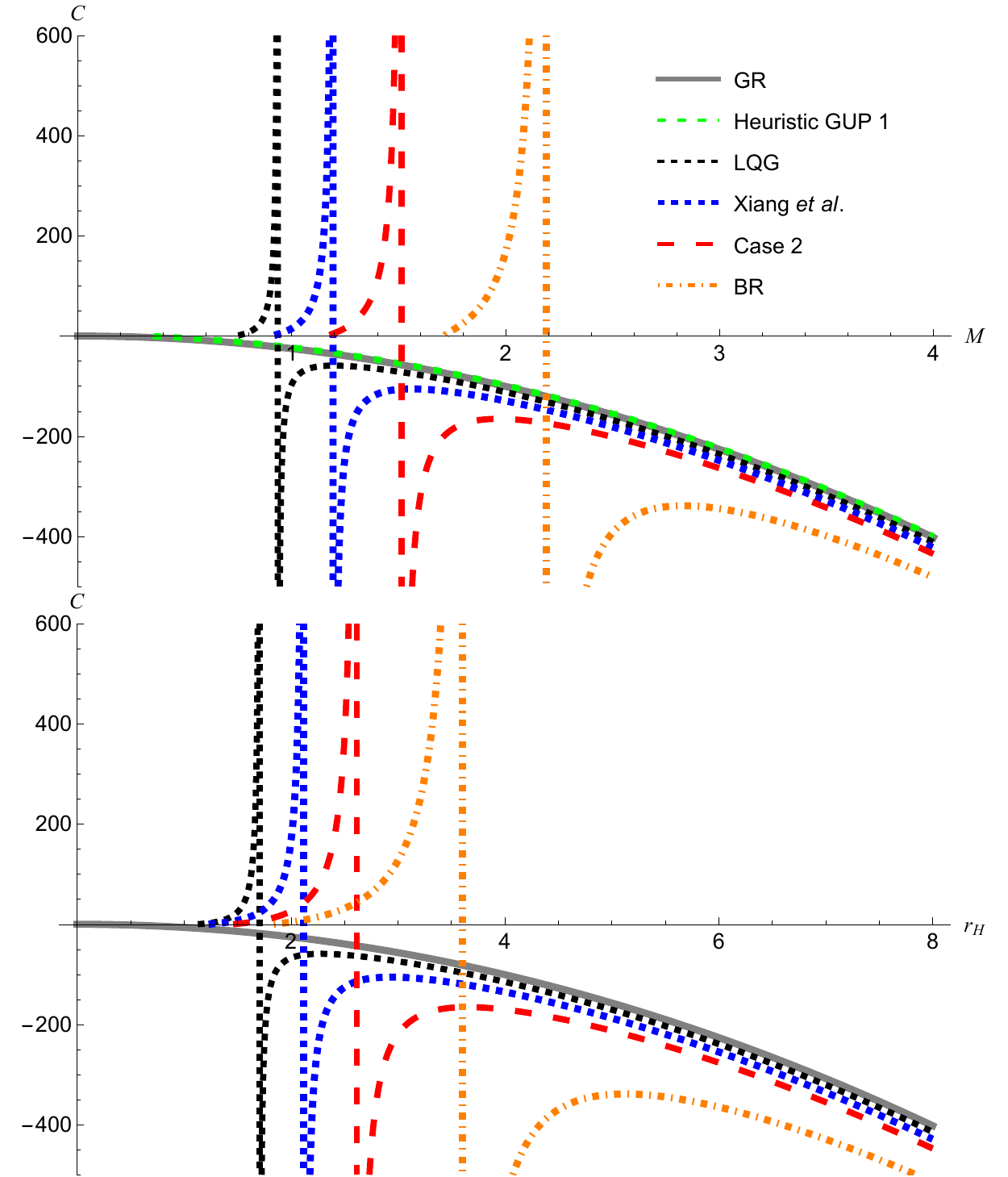}
    \caption{Heat capacity as a function of mass and horizon radius.}
    \label{fig:5}       
\end{figure*}

\begin{figure*}[ht!]
    \centering
    \includegraphics[scale=0.5]{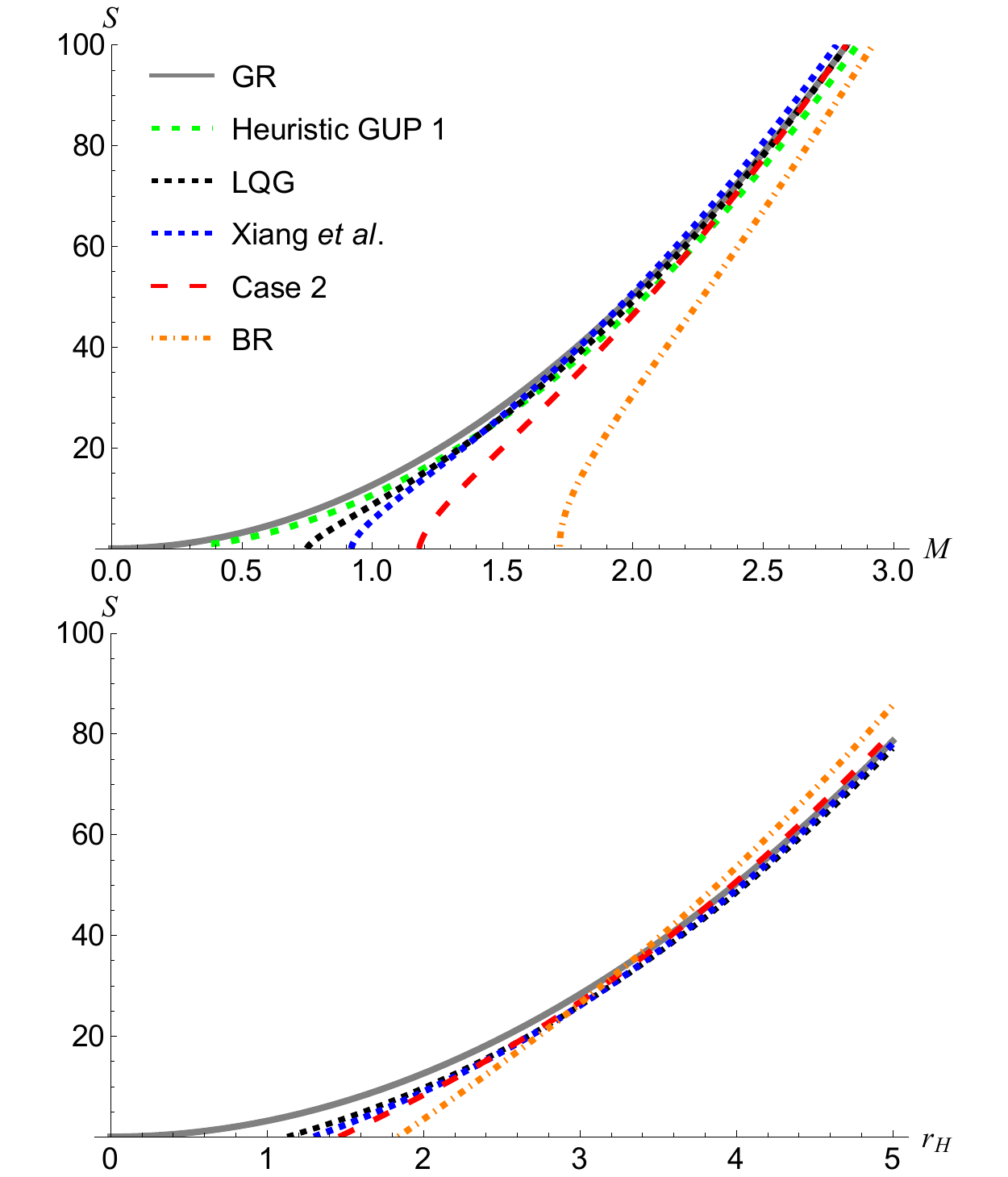}
    \caption{Entropy as a function of mass and horizon radius.}
    \label{fig:6}       
\end{figure*}

The identification of black holes as thermodynamic objects leads to exploring their similarities with other thermodynamic systems, such as phase transitions. The author in Ref. \cite{Roychowdhury:2013klt} has comprehensively studied the black hole phase transition in various cases. The phase transition is a distinctive feature of the quantum-corrected BH \cite{Shahjalal:2018hid,Kim:2012cma}, and is evident in its heat capacity, which indicates the thermodynamic stability region of the BH. The heat capacity can be written as 
\begin{align}
    C=\frac{\mathrm{d}M}{\mathrm{d}T}. \label{Heat capacity}
\end{align}
Inserting the Hawking temperature given by Eqs.~\eqref{Hawking temperature Ong} and \eqref{Our Hawking temperature}-\eqref{Lewandowski Hawking temperature} into Eq.~\eqref{Heat capacity} and Eqs.~\eqref{Our black hole mass}-\eqref{Lewandowski black hole mass}, we obtain the evolution of the BH heat capacity as the mass increases (top panel from Fig.~\ref{fig:5}) and as the radius of the horizon increases (bottom panel). The divergence point shows the phase transition, which appears once in each curve. The relatively stable BH showed positive heat capacity and occurred in a small mass regime. In other words, our results show that the larger BH behaves less thermodynamically stable than the smaller one. This behavior shares a different result with the GUP effect in the Rutz-Schwarzschild BH \cite{Fu:2021zrd} that predicts two phase transitions, produces two stable BH regimes (in small and large BHs), and is unstable in the intermediate mass regime. Fig.~\ref{fig:5} shows that all corrections reduce to GR as standard heat capacity $-2\pi r_H^2$ in the large-mass regime. We want to highlight that the minimum mass in the BR metric was significantly influenced by $\hat{\gamma}$, as we can see in Eq.~\eqref{Bonanno-Reuter black hole mass}. If we insert $\hat{\gamma}=9/2$ as proposed in their work, the minimum mass becomes larger and the coincidence of its heat capacity with GR will occur in a much larger mass regime. From our previous results, as we already stated in Table \ref{table 1}, if we use their $\tilde{\omega}$, we can get the range of the corresponding $\hat{\gamma}$. Inserting the lower bound $\hat{\gamma}\approx 0.188$, we obtain the result as in Fig.~\ref{fig:5}. This value is also used in the other thermodynamic and shadow properties to maintain consistency. We also want to highlight that the heat capacity of the standard heuristic GUP, as shown by the dashed green line, did not exhibit a phase transition. 

One of the thermodynamic properties we also discuss is entropy, which is defined as
\begin{align}
    S&=4\pi \int{M \mathrm{d}r_H}, \label{S(rH)} \\
    S&=\int{\frac{\mathrm{d}M}{T}}, \label{S(m)}
\end{align}
where it can be used freely, depending on whether we want to plot against mass or horizon radius. In the BH context, we expect that the presence of scale-dependent Newton constants in our metric modifies the horizon radius, thereby altering the thermodynamic properties. Inserting Eq.~\eqref{Our black hole mass} into Eq.~\eqref{S(rH)}, we obtain the entropy as
\begin{align}
    S_{TW}=\frac{\pi r_H^2}{\ell_p^2}+\frac{1}{2}\pi(2+\hat{\gamma})\beta_0\ln{\left[2r_H^2-\ell_p^2\beta_0\hat{\gamma}\right]}. \label{our entropy}
\end{align}
 Inserting Eqs.~\eqref{Xiang black hole mass} and \eqref{Bonanno-Reuter black hole mass} into Eq.~\eqref{S(rH)}, we obtained the following entropy:
\begin{align}
    S_{Xi}&=\frac{\pi r_H^2}{\ell_p^2}+\pi\beta_0\ln{\left[r_H^2-\ell_p^2\beta_0\right]}, \label{Xiang entropy} \\
    S_{BR}&=\frac{\pi r_H^2}{\ell_p^2}+\frac{1}{2}\pi(2+\hat{\gamma})\tilde{\omega}\ln{\left[2r_H^2-\ell_p^2\tilde{\omega}\hat{\gamma}\right]}. \label{Bonanno-Reuter Entropy}
\end{align}

For the sake of effectiveness, we do not present the full expression of the analytical entropy of \cite{Lewandowski:2022zce} due to its length. Its leading term is the Bekenstein-Hawking entropy of GR with logarithmic corrections. The coefficient of the logarithmic term depends on the specific model, while only the argument of the logarithm changes for smaller $r_H$. For example, in some models, the argument involves higher-order curvature terms or additional coupling constants. In particular, the coefficient remains positive in all cases; thus, for large $r_H$, the entropy approaches the GR result. For small $r_H$, although the overall trend remains similar, specific features and subleading terms can differ between models due to their different assumptions and parameter choices. The universality of the logarithmic correction is addressed in Ref.~\cite{Das:2001ic} and the citations therein.

Figure~\ref{fig:6} shows that our proposed GUP metric provides a nonzero mass when entropy reaches zero, indicating a remnant mass. Notably, in Hawking temperature, heat capacity, and entropy, our metrics yield the same remnant mass. This result differs from rainbow gravity in~\cite{Ali:2014xqa}, where only Hawking temperature signals the remnant mass, and the minimum masses in entropy and heat capacity represent the information contained in the remnant of BH.

The standard heuristic GUP predicts a minimum mass proportional to \(\sqrt{\beta_0}\). From the entropy, this prediction appears as a nonzero mass when the entropy vanishes. However, this heuristic GUP does not show a remnant mass in the Hawking temperature, and thus there is no phase transition in the heat capacity. In contrast, our proposed GUP-effective metric predicts the remnant mass through both the Hawking temperature and entropy, and it demonstrates a phase transition in heat capacity. Therefore, we infer that our effective metric provides a physically more reasonable description of a phenomenologically quantum-gravity-motivated black hole thermodynamics than the heuristic approach.

\section{Black-hole shadow and optical appearance}
\label{sec:SP}
In this section, we investigate the optical properties of the BH models. These include the photon sphere radius and its resulting shadows. We derive these properties by studying geodesic equations in spacetime, particularly by analyzing the photon effective potential. An exciting aspect of BH shadows arises when they are generated in the presence of surrounding accretion disks. We achieve this by a ray-tracing procedure from a distant observer to the BH, while modeling an accretion disk around it. Before we present the images, we describe the accretion disk models and their properties. We then discuss and compare the image features of each BH model—especially in their minimum mass configurations—and relate them to their corresponding shadow radii. In the discussion, we present two comparisons. Specifically, for each case listed in Table~\ref{table 2}, we conducted one comparison for each black hole (BH) model: Loop Quantum Gravity (LQG)~\cite{Lewandowski:2022zce}, the Xiang \textit{et al.} model~\cite{Li:2016yfd}, the Bonanno-Reuter (BR) model~\cite{Bonanno:2000ep}, and Case 2 of this work.

\subsection{Photon geodesics}
\begin{figure*}[htbp!]
    \centering
    \includegraphics[scale=0.65]{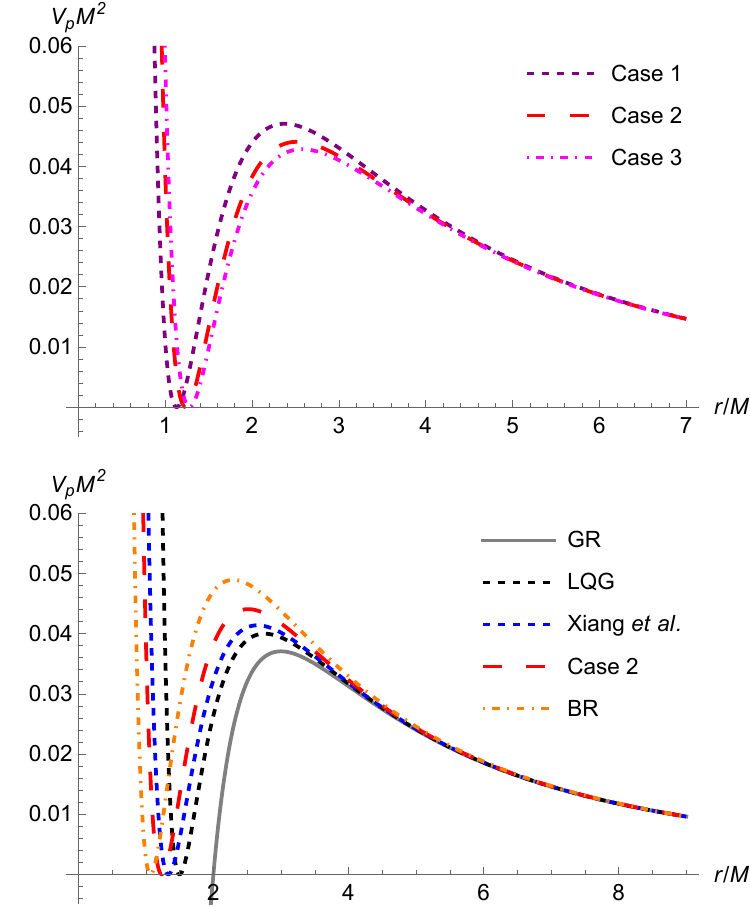}
    \caption{Dimensionless effective potential for photon geodesic in the minimum mass configuration for each BH models.}
    \label{fig. photon effective potential}
\end{figure*}

We start from the geodesic equation given by
\begin{equation}
\frac{\mathrm{d}\dot{x}^{\mu}}{\mathrm{d}\tau} + \Gamma^{\mu}_{\alpha\beta} \frac{\mathrm{d}x^{\alpha}}{\mathrm{d}\tau} \frac{\mathrm{d}x^{\beta}}{\mathrm{d}\tau} = 0,
\end{equation}
where $\tau$ is an affine parameter, with $x^{\mu} = (t, r, \theta, \phi)$ and $\dot{x}^{\mu} = \mathrm{d}x^{\mu}/\mathrm{d}\tau$.
By solving the geodesic equation and restricting the motion to the equatorial plane ($\theta = \pi/2$, $\dot{\theta} = 0$), we obtain two conserved quantities:
\begin{equation}
A(r)\dot{t} = E, \qquad r^2\dot{\phi} = L,
\label{eq. const of motion}
\end{equation}
where $E$ and $L$ are interpreted as the energy and the angular momentum of the test particle, respectively. Using the null geodesic condition,
\begin{equation}
-A(r)\dot{t}^2 + A(r)^{-1} \dot{r}^2 + r^2\sin^2\theta \dot{\phi}^2 = 0,
\label{eq. null condition}
\end{equation}
we eventually arrive at the geodesic equation for a light ray:
\begin{equation}
\frac{\mathrm{d}\phi}{\mathrm{d}r} = \frac{1}{r^2} \left[b^{-2} - V_p(r)\right]^{-1/2}, \qquad V_p(r) = \frac{A(r)}{r^2},
\label{radial geodesic equation}
\end{equation}
or equivalently,
\begin{equation}
\frac{\mathrm{d}r}{\mathrm{d}\phi} = \pm r^2 \left[b^{-2} - V_p(r)\right]^{1/2},
\end{equation}
where $V_p(r)$ is the photon effective potential and $b = L/E$ is the impact parameter. Fig.~\ref{fig. photon effective potential} $ V_p M^2$ shows the effective potential $V_p M^2$ of dimensionless photons for each model, in its extremal or minimum mass configuration. Generally, the profile of $V_p M^2$ is not significantly influenced by the values of $\beta_0$ and $\hat{\gamma}$; all quantum BH models exhibit similar trends in their $V_p M^2$ profiles. The effective potential displays a clear pattern: the height of the local maximum is inversely related to its location—higher local maxima occur slightly closer to the center of the BH. This points to a proportional relationship between the radius of the photon sphere and the BH’s shadow radius, since the local maximum marks the (unstable) radius of the photon sphere, while the inverse of the height of the effective potential in the photon sphere is related to the shadow radius. We analyze both quantities in the following discussion.

\begin{figure*}[htbp!]
    \centering
    \includegraphics[scale=0.65]{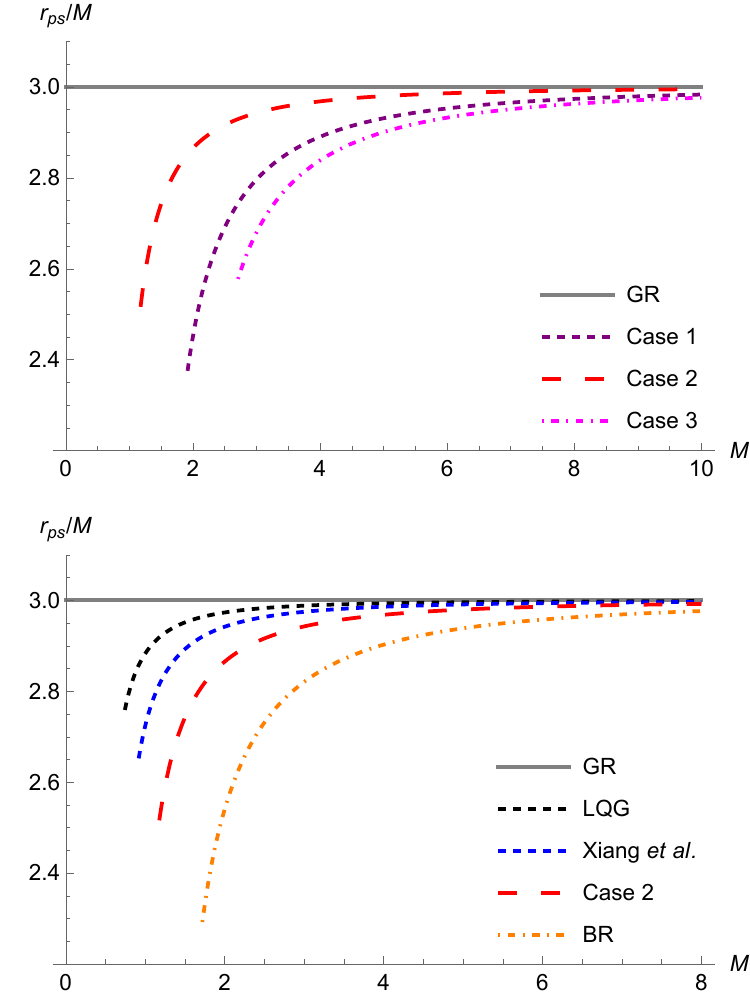}
    \caption{The radius of the photon sphere $r_{ps} $ as a function of BH's mass. Each of the curve started from the minimum mass of the corresponding BH.}
    \label{fig. photon sphere radius}
\end{figure*}

Within the photon geodesic equation, one can determine the corresponding photon sphere radius $r_{ps}^\pm$. This is obtained by applying conditions $\mathrm{d}r/\mathrm{d}\phi = \mathrm{d}^2r/\mathrm{d}\phi^2 = 0$ at $r = r_{ps}^\pm$, which leads to the extremum condition of the effective potential:
\begin{equation}
\left.\frac{\mathrm{d}V_p}{\mathrm{d}r}\right|_{r=r_{ps}^\pm}=0,
\end{equation}
where $\pm$ denotes the outer ($+$) unstable photon sphere and the inner ($-$) stable photon sphere. In the context of BH spacetime, only the outer unstable photon sphere is physically relevant, as the inner photon sphere always lies within (or at) the event horizon and therefore is unobservable to an external observer. Hence, we focus solely on $r_{ps}^+$ and drop the $+$ sign for convenience. We compute the photon sphere radius numerically for each spacetime configuration discussed in the previous section, as shown in Fig.~\ref{fig. photon sphere radius}.

Our investigation reveals that, in our proposed BH model, the chosen $\beta_0$ does not affect the photon sphere radius in the minimum mass configuration. The parameter $\beta_0$ influences only the minimum mass value: $M_{\min}$ increases as $\beta_0$ increases. Instead, the photon sphere radius in their minimum mass is fully controlled by the parameter $\hat{\gamma}$; a larger $\hat{\gamma}$ yields both a larger photon sphere radius and a larger minimum mass. This behavior is evident in the representative parameter sets considered in our study (top panel of Fig.~\ref{fig. photon sphere radius}): Case 1, with the smallest $\hat{\gamma}$, results in the smallest photon sphere radius, while Case 3, with the largest $\hat{\gamma}$, produces the largest. Hence, for minimum mass BHs, the minimum length scale does not affect the photon sphere radius, which is determined entirely by the new parameter $\hat{\gamma}$ appearing in the $(\Delta p)^2$ relation. Moreover, $\beta_0$ governs the rate at which the photon sphere radius converges to the Schwarzschild value $r_{ps} = 3M$ at large mass, where convergence is faster for smaller $\beta_0$.

Comparing the BH models, one can see that the LQG BH has a photon sphere radius that deviates slightly from the Schwarzschild case and exhibits a faster convergence rate at large mass. In contrast, the BR BH model shows the slowest convergence and the smallest photon sphere radius, despite having the largest minimum mass. Our Case 2 model lies between these two models.

\subsection{Shadow radius}
\label{sec. shadow radius}
\begin{figure*}[htbp!]
    \centering
    \includegraphics[scale=0.65]{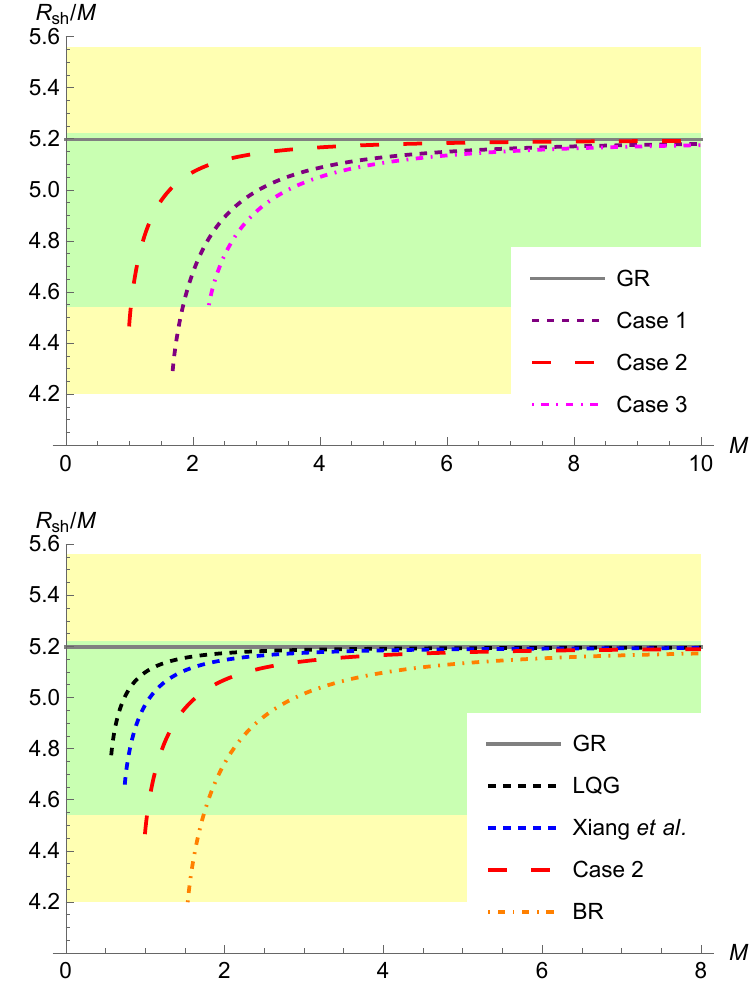}
    \caption{Shadow radius as a function of BH's mass. Each of the curve started from the minimum mass of the corresponding BH. The yellow and green region are the $2\sigma$ and $1\sigma$ Sgr A* data constraint, respectively.}
    \label{fig. shadow radius}
\end{figure*}

The shadow radius can be obtained by determining the photon’s critical impact parameter $b_c$ corresponding to the formation of a photon sphere. This is done by finding the value of $b$ for which the radial motion in Eq.~\eqref{radial geodesic equation} vanishes at the photon sphere, yielding
\begin{equation}
b_c = \sqrt{\frac{1}{V_p(r_{ps})}}.
\end{equation}
Assuming the observer is located far away, effectively at infinity, the shadow radius is equal to the critical impact parameter, \textit{i.e.}, $R_{sh} = b_c$ (for details, see, e.g., Ref. \cite{Rohim:2025gxo}). In Fig.~\ref{fig. shadow radius}, one can observe that the shadow radius exhibits a behavior similar to that of the photon sphere radius (cf. Fig.~\ref{fig. photon sphere radius}).

The analysis of the BH shadow commonly employs the EHT constraint on the shadow radius of the supermassive Sgr A* BH. This constraint is given by~\cite{Vagnozzi:2022moj}
\begin{align}
4.22 < R_{sh}/M < 5.56 \qquad &(2\sigma)\\
4.58 < R_{sh}/M < 5.21 \qquad &(1\sigma)
\end{align}
Although this constraint may be less relevant for low-mass BHs (as deviations from Schwarzschild only appear near the minimum mass), it still provides an interesting perspective on the possibility that the Sgr A* BH is in a minimum mass configuration. Our results indicate that the constraint $1\sigma$ Sgr A * accommodates the minimum mass configurations of Case 3, LQG, and Xiang \textit{et al.} BHs. The minimum mass configurations of the other models fall within the $2\sigma$ range, with the BR BH model producing a slightly smaller shadow radius than the lower bound. Therefore, we conclude that all models considered in our study are in agreement with the Sgr A* data\textemdash at least within the $2\sigma$ range\textemdash suggesting that Sgr A* could, in principle (albeit unlikely), be described by a minimum mass BH in one of the models discussed.

It is important to note that for supermassive or macroscopic BHs, the impact of quantum effects is expected to be very small because the ratio between Planck length and the radius of BH is extremely small. However, in this subsection, we only check whether the IR part of the metric of this effective model is still acceptable and whether there is still a small signature of quantum effects traced in this kind of BH.  The effect indeed becomes observationally relevant in micro black holes. Unfortunately, observational results for the shadows of this kind of BH do not exist until now.

\subsection{Thin accretion disk}
We assume an optically and geometrically thin accretion disk with monochromatic emission. For simplicity, we consider an accretion disk emission profile $I_e(r)$, which is related to the matter density and temperature of the disk. We adopt the emission profile known as the Gralla-Lupsasca-Marrone (GLM) model~\cite{Gralla:2020srx},
\begin{equation}
I_e(r) = \frac{\exp\left\{-\frac{1}{2}\left[\gamma + \operatorname{arcsinh}{\left(\frac{r-\mu}{\sigma}\right)}\right]^2\right\}}{\sqrt{\left(r-\mu\right)^2+\sigma^2}},
\label{eq. intensity profile glm}
\end{equation}
where $\gamma$, $\mu$, and $\sigma$ are parameters that determine the shape of the accretion disk. In contrast to exponential cut-off models (e.g., Refs.~\cite{Meng:2023htc,Zeng:2023fqy,Guo:2022iiy}), the GLM profile provides a continuous intensity function with adjustable smoothness, governed by the parameters $\sigma$ and $\gamma$. This model has been shown to closely align with observational predictions of intensity profiles for astrophysical accretion disks, as derived from general relativistic magnetohydrodynamics simulations~\cite{Vincent:2022fwj}, and has been applied in various studies of the appearance of BHs and ultracompact objects~\cite{Rosa:2023hfm,daSilva:2023jxa, Fauzi:2024nta}.

In this study, we adopt an accretion profile characterized by $\gamma = -2$, $\mu = R_{ISCO}$, and $\sigma = M/4$, where $R_{ISCO}$ is the radius of the innermost stable circular orbit (ISCO) for a massive particle. This accretion disk assumes that the emission peaks near the ISCO and vanishes beyond it. The value of $R_{ISCO}$, for a spacetime geometry given in the form of Eq.~\eqref{GUP-Modified SCH}, can be calculated by numerically solving~\cite{Gao:2023mjb}
\begin{equation}
\left.3A(r)A'(r) - 2rA'(r)^2 + rA(r)A''(r)\right|_{r=R_{ISCO}} = 0.
\end{equation}

We take into account the gravitational redshift effect of the accretion disk as seen by the observer. We assume a static accretion disk, whose four-velocity is expressed as
\begin{equation}
u^{\mu}_e = (u^t,0,0,0).
\label{eq. four velocity accretion}
\end{equation}
The $u^t$ component can be obtained from the normalization condition for timelike particles, yielding
\begin{equation}
u^t_e(r) = \frac{1}{\sqrt{A(r)}}.
\end{equation}

To include redshift effects, the observed photon frequency $\nu_o$ is related to the emitted frequency $\nu_e$ by the energy correction factor $\tilde{g}$, written as~\cite{Ozel:2021ayr,M:2022pex,Kumaran:2023brp}
\begin{equation}
\nu_o = \tilde{g} \nu_e, \qquad
\tilde{g} = \frac{-k_{\nu} u^{\nu}_o}{-k_{\mu} u^{\mu}_e},
\label{eq. energy corr f}
\end{equation}
where $k{\mu} = g_{\mu\nu}k^{\nu}$ is the covariant photon four-momentum and $u^{\nu}_o$ is the four-velocity of the observer. Assuming a static observer is located far away in flat space-time, we have $u^{\nu}_o = (1,0,0,0)$, and therefore
\begin{equation}
\tilde{g} = \frac{1}{u^t_e} = \sqrt{A(r)}.
\end{equation}

\begin{figure*}[htbp!]
    \centering
    \includegraphics[scale=0.62]{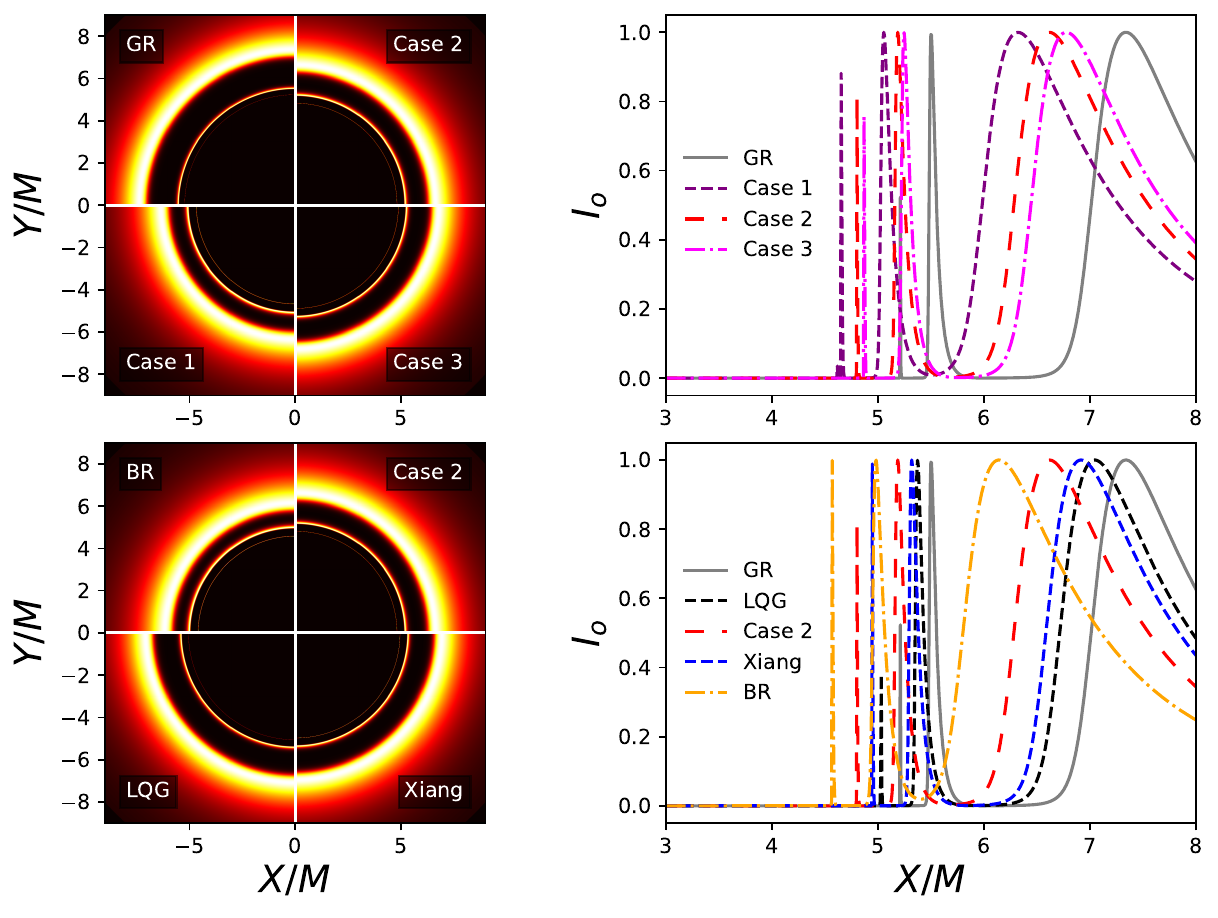}
    \caption{(Left) Comparison of the BH's appearance surrounded by accretion disk in their minimum mass configuration. (Right) Intensity cross section of the BH images.}
    \label{fig. shadow images}
\end{figure*}

The specific intensity $I^\nu$ is related to the frequency via the Lorentz-invariant quantity $I^\nu/\nu^3$, which implies $I^\nu_e/\nu_e^3 = I^\nu_o/\nu_o^3$~\cite{M:2022pex}.
Integrating over all frequencies, the total observed intensity becomes
\begin{equation}
I_o(r) = \int I^\nu_o(r) \mathrm{d}\nu_o = A(r)^2 I_e(r).
\end{equation}
It is important to note that as long as we consider macroscopic BH such as Sgr A* that we used in this work, we can use the GLM accretion disk emission profile. However, for micro BHs where quantum effects are actually important, we cannot use such a classical ISCO profile.

\subsection{Black-hole appearance}
We generate the images of the BH models using a ray-tracing procedure based on the backward integration method. Specifically, we numerically integrate the photon geodesics from Eq.~\eqref{radial geodesic equation}, starting from an observer at numerical infinity $r_o$ and tracing them toward the BH. The ray-tracing procedure used in this study follows the same approach as in Ref.~\cite{Fauzi:2024nta} for axial observation, with the observer placed at $r_o = 500M$.

Considering the shadow radius results from Sect.~\ref{sec. shadow radius}.  ~we analyze the appearance of BHs surrounded by accretion disks in their minimum-mass configurations to highlight the most significant differences between the BH models. The generated images and their corresponding intensity cross sections are shown in Fig.~\ref{fig. shadow images}.

All BH images share a common feature: two (inner) intensity peaks\textemdash dubbed as the \textit{ photon ring}\textemdash produced by extreme light deflection near the BH. The shadow boundaries of the BHs are seen as the regions enclosed by the innermost thin photon rings. Our resulting images agree with our shadow calculation in Fig.~\ref{fig. shadow radius}. A particularly new insight from these images is the distance between the intensity peak of the direct emission from the accretion disk and that of the outermost photon ring. One can qualitatively see that the peak distance decreases with decreasing shadow radius, with Schwarzschild and BR BHs showing the most significant and smallest distances, respectively. Furthermore, we find that a smaller shadow radius corresponds to a smaller ISCO radius. Hence, this peak arises from differences in the ISCO and photon sphere radii, which alter the accretion disk and the BH's image characteristics.

It should be noted that our image analysis was conducted in the BH minimum mass configuration, which may be extremely small. If one adopts the actual value of $\ell_p$, the minimum mass of a BH would be on the order of the Planck scale. This behavior renders the accretion disk's features irrelevant, since we treat it as a circularly orbiting stream of massive particles\textemdash an assumption that breaks down at the Planck scale. In the higher mass regime where $M \gg \ell_p$, the shadow observables of all BH models converge to those predicted by GR, making them indistinguishable from classical GR.

This problem challenges the feasibility of probing quantum gravity through BH optical observables and may require novel scenarios to push the limits of quantum gravity theories. For instance, several studies (e.g., Refs.~\cite{Eichhorn:2022bbn, Fauzi:2025ldu}) have explored the observational possibility of destroying the BH horizon by overspinning the object\textemdash particularly in the case of regular BHs, which are not constrained by the weak cosmic censorship conjecture. Such scenarios can lead to significant differences in the image features, even when the (quantum) corrections in the large-mass regime remain small.

To this end, it is also interesting to note that the peak distance can be a potential observable for determining the corresponding BH spacetime. It may be observed by future instruments, which might resolve the inner photon ring feature in the BH image. However, several aspects could lead to the same behavior: BH spin and charge. It has been shown that these two quantities also affect the ISCO and the BH's shadow radius, which, within GR itself, decrease both quantities as the spin and/or charge increase \cite{Fauzi:2025ldu,Abdujabbarov:2016hnw,Meng:2024puu,Xavier:2020egv}. Thus, to truly determine the spacetime structure around the BH, one might need a comprehensive analysis of other observables, such as massive particle orbits (e.g., stars) and gravitational lensing.

\section{Connection quadratic GUP with $f(R)$ gravity}\label{sec:FR}
In the previous sections, we discussed how to build an effective GUP metric while preserving the minimal-length imprint through thermodynamic and shadow observables. However, it is still unusual to consider GUP as a framework for understanding quantum effects in gravity since there is no robust guidance to obtain GUP-effective gravitational Lagrangians. Lately, \cite{DAgostino:2025axy} proposed a way to tackle this lack utilizing the Wald entropy formula to obtain the GUP effective Lagrangian. Applying their proposal to our investigation, we define a deformation encoded by the quadratic GUP at the action level by calculating its effective Lagrangian density. 

First, we want to identify the GUP effect on generalized action -- a general prescription that we particularly use in modified gravity, which takes the form:
\begin{equation}
    \mathcal{S}_g=\frac{1}{16\pi}\int_\mathcal{H} \mathrm{d}^4x\sqrt{-g}f(R). \label{generalized action}
\end{equation}
Using the formulation of Wald entropy in $f(R)$ gravity \cite{DAgostino:2024sgm}, and considering the horizon area as $4\pi r_H^2$, we obtain
\begin{equation}
    f_R(R_H)=4R_HS_{TW}\ell_p^2,
\end{equation}
where $f_R(R_H)$ is defined as the derivative of $f(R_H)$ with respect to the Ricci scalar $R$, and $R_H$ is the Ricci scalar at the horizon radius. The Ricci scalar is a geometric scalar invariant, so $R_H\leftrightarrow R$ and $f(R_H)\leftrightarrow f(R)$. The entropy of the quadratic GUP $S_{TW}$ takes the form \eqref{our entropy}, where the horizon radius $r_H$ can be transformed to $R_H$ by the following relation:
\begin{equation}
    r_H=\frac{1}{\sqrt{4\pi R_H}}.
\end{equation}
Therefore, $R_H$ can once again be transformed into $R$. The derivative of $f(R)$ can now be obtained as follows:
\begin{equation}
    \frac{\mathrm{d}f(R)}{\mathrm{d}R}=1+2\ell_p^2\pi R(2+\hat{\gamma})\mathrm{ln}\bigg[\frac{1}{2\pi R}-\ell_p^2\beta_0\hat{\gamma}\biggl].\label{our df(R)}
\end{equation}
By integrating Eq.~\eqref{our df(R)}, we obtain the corresponding $f(R)$ form of the quadratic GUP, which takes the explicit form as 
\begin{align}
    f(R)&=\ell_p^2\pi R^2(2+\hat{\gamma})\mathrm{ln}\bigg[\frac{1}{2\pi R}-\ell_p^2\beta_0\hat{\gamma}\biggl]\nonumber\\
    &+\frac{2\ell_p^2\pi R\beta_0\hat{\gamma}(-2-\hat{\gamma}+
    2\beta_0\hat{\gamma})-(2+\hat{\gamma})\mathrm{ln}[1-2\ell_p^2\pi R\beta_0\hat{\gamma}]}{4\ell_p^2\pi\beta_0^2\hat{\gamma}^2} + \mathcal{C}, \label{GUP Lagrangian}
\end{align}
Here, $\mathcal{C}$ denotes the integration constant. This result for Eq.~\eqref{GUP Lagrangian} completes the expression for the generalized action in Eq.~\eqref{generalized action}. It explicitly illustrates how the quadratic GUP entropy may translate into the gravitational action in $f(R)$ form\cite{DAgostino:2025axy}.
\section{Conclusions}
\label{sec:Conclu}
In this work, we study the momentum-dependent $\Delta p$ metric for the quadratic GUP (KMM model), based on the gravity-induced phase shift (COW) and the Einstein-Bohr Gedanken  for the photon box experiments. Instead of the $\Delta p(r)$ relation predicted by the tidal force~\cite{Li:2016yfd}, we use an interpolating $\Delta p(r)$ relation inspired by the effective distance from RG black hole studies~\cite{Bonanno:2000ep}. This effective parameterization captures the behavior of $\Delta p(r)$ at both small (UV) and large (IR) distances. Using this function, we find the same effective Newton constant structure as in RG theory~\cite{Bonanno:2000ep}, making the metric regular in the center. Since our metric resembles that in~\cite{Bonanno:2000ep}, we constrain the GUP parameter $\beta_0$ via $\beta_0\zeta^2\equiv\tilde{\omega}$, using quantum corrections to the Newtonian potential~\cite{Bonanno:2000ep}. In addition, the similarity between our metric and the LQG metric~\cite{Lewandowski:2022zce} allows us to obtain $\hat{\gamma}\equiv8\sqrt{3}\pi\gamma/\beta_0$, which RG BH~\cite{Bonanno:2000ep} lacks. The minimum uncertainty length exists only if $\beta_0\equiv\tilde{\omega}>0$, so we use only positive quantum corrections from the literature~\cite{Bonanno:2000ep,Donoghue:1993eb,Bjerrum-Bohr:2002fji} to constrain $\beta_0$. Given the uncertainty in the constraints (Table~\ref{table 1}), we consider three cases in which both parameters vary to study the thermodynamics, shadows, and appearance of the BH. Each combination of $\beta_0$ and $\hat{\gamma}$ is in Table~\ref{table 2}. Finally, the BH's deformed metric solves the quantum Raychaudhuri equation, and the sign of $\beta_0$ aligns with~\cite{Jana:2025hgv}. We also found that our effective metric improved the issue that existed in the heuristic GUP metric  in Eq.~\eqref{metric Ong} in the UV region.

In BH thermodynamics, we study the Hawking temperature, heat capacity, and entropy from the effective metric derived from our proposed GUP. We compare our results with the standard heuristic approach and other quantum BH predictions. Unlike the heuristic approach, our metric gives the same remnant mass for both Hawking temperature and entropy. Our results show BH thermodynamics trends similar to those of RG and LQG BHs. A finite minimum mass occurs when the Hawking temperature and entropy reach zero. Both RG and LQG approaches exhibit a divergence in heat capacity that signals a phase transition, a prediction our metric also makes. The heuristic approach does not show this phase transition. The metric from~\cite{Li:2016yfd} is the same as ours for $\hat{\gamma}$ = 2 and $\zeta$ = 1, but only when $r\ll\hat{\gamma}$GM. The parameters $\hat{\gamma}$ and $\beta_0$ shift the minimum or remnant mass. In contrast, the LQG metric in~\cite{Lewandowski:2022zce} returns to GR for small Barbero-Immirzi $\gamma$, while our metric retains its correction even for small $\gamma$. 

We study the shadow and optical properties of our model and compare them to those of RG and LQG, finding similar trends. For minimum mass BHs, the photon sphere and shadow radius are not affected by the GUP parameter $\beta_0$; they are fully controlled by $\hat{\gamma}$. Although $\beta_0$ does not affect these features at minimum mass, it does affect how fast the model approaches the Schwarzschild photon sphere and shadow radius at high mass. Our model offers an alternative to the LQG and RG BHs, possibly better matching the Sgr A* data for the minimum mass case compared to the RG BH in Ref.~\cite{Bonanno:2000ep}. In case three, where $\beta_0=1.5$ and $\hat{\gamma}=9/2$, our model fits within the 1$\sigma$ Sgr A* constraint along with the models from \cite{Lewandowski:2022zce} and \cite{Li:2016yfd}.

In summary, our scale-dependent GUP metric demonstrates internal consistency of the proposed phenomenological framework and reproduces several commonly discussed quantum-gravity-inspired black hole features, such as minimal length effects, regularized geometry, and possible remnants. The construction of this metric is based on the gravity-induced phase shift (the COW experiment) and the Einstein-Bohr photon box  Gedanken experiment. We augment it with the interpolating function $\Delta p(r)$ inspired by Ref.~\cite{Bonanno:2000ep}. Compared with the heuristic approach and the GUP metric from \cite{Ong:2023jkp}, our results match better in the quantum BH features. The heuristic GUP does not predict remnants with zero entropy and zero Hawking temperature because it does not account for UV behavior. Our metric allows us to study GUP effects in gravitationally bound objects, including horizonless ones such as neutron stars and white dwarfs. The procedure in Ref. \cite{DAgostino:2025axy} based on the GUP effective metric and the $f(R)$ gravity correspondence leads us to a future route toward a more fundamental description of GUP.


\section*{Acknowledgments}
Thanks to Byon Nugraha Jayawiguna for the fruitful discussion. We also want to express our sincere gratitude to the Ministry of Higher Education, Science, and Technology of the Republic of Indonesia for their support through the PMDSU program scholarship and grant fund, with the grant numbers 070/C3/DT.05.00/PL/2025 and PKS-578/UN2.RST/HKP.05.00/2025.

\end{document}